\documentclass[iop,apj,numberedappendix,appendixfloats]{emulateapj}
\usepackage{color}
\usepackage{amsmath}

\newcommand{\solar}{$_{\odot}$}
\newcommand{\tco}{$^{12}$CO}
\newcommand{\ttco}{$^{13}$CO}
\newcommand{\ceto}{C$^{18}$O}
\newcommand{\hcop}{HCO$^+$}

\newcommand{\nnh}{N$_2$H$^+$}
\newcommand{\nht}{NH$_3$}

\newcommand{\joz}{$J$=1$\rightarrow$0}

\newcommand{\kms}{\,km\,s$^{-1}$}
\newcommand{\degree}{$^{\circ}$}
\newcommand{\fdeg}{$^{\circ}$\hspace{-1mm}.}

\newcommand{\nf}{$\nu_N$}

\newcommand{\tsys}{$T_{\rm sys}$}
\newcommand{\tex}{$T_{\rm ex}$}
\newcommand{\tmb}{$T_{\rm mb}$}
\newcommand{\tbg}{$T_{\rm bg}$}
\newcommand{\vlsr}{$V_{\rm LSR}$}
\newcommand{\brg}{Br-$\gamma$}
\newcommand{\htwo}{H$_2$}

\newcommand{\nhtwo}{$N_{\rm H_2}$}
\newcommand{\ico}{$I_{\rm ^{12}CO}$}
\newcommand{\lco}{$L_{\rm CO}$}
\newcommand{\xco}{$X_{\rm CO}$}
\newcommand{\nco}{$N_{\rm CO}$}

\newcommand{\tbtco}{$T_{\rm ^{12}CO}$}
\newcommand{\ihcop}{$I_{\rm HCO^+}$}
\newcommand{\xhcop}{$X_{\rm HCO^+}$}
\newcommand{\thcop}{$T_{\rm HCO^+}$}

\def\lapp{\ifmmode\stackrel{<}{_{\sim}}\else$\stackrel{<}{_{\sim}}$\fi}
\def\gapp{\ifmmode\stackrel{>}{_{\sim}}\else$\stackrel{>}{_{\sim}}$\fi}

\shorttitle{CHaMP III.  \tco\ in Dense Clump Envelopes}
\shortauthors{Barnes et al.}

\begin{document}

\title{The Galactic Census of High- and Medium-mass Protostars.  III \tco\ Maps and \\
    Physical Properties of Dense Clump Envelopes and their Embedding GMCs}

\author{Peter J. Barnes\altaffilmark{1,2}, Audra K. Hernandez\altaffilmark{3}, Stefan N. O'Dougherty\altaffilmark{4}, \\
William J. Schap III\altaffilmark{1}, and Erik Muller\altaffilmark{5}
}
\email{pjb@ufl.edu}

\altaffiltext{1}{Astronomy Department, University of Florida, P.O. Box 112055, Gainesville, FL 32611, USA}
\altaffiltext{2}{School of Science and Technology, University of New England, Armidale NSW 2351, Australia}
\altaffiltext{3}{Astronomy Department, University of Wisconsin, 475 North Charter St., Madison, WI 53706, USA}
\altaffiltext{4}{College of Optical Sciences, University of Arizona, 1630 E. University Blvd., P.O. Box 210094, Tucson, AZ 85721, USA}
\altaffiltext{5}{National Astronomical Observatory of Japan, Chile Observatory, 2-21-1 Osawa, Mitaka, Tokyo 181-8588, Japan}

\begin{abstract}
We report the second complete molecular line data release from the {\em Census of High- and Medium-mass Protostars} (CHaMP), a large-scale, unbiased, uniform mapping survey at sub-parsec resolution, of mm-wave line emission from 303 massive, dense molecular clumps in the Milky Way.  This release is for all \tco\ \joz\ emission associated with the dense gas, the first from Phase II of the survey, %
which includes \tco, \ttco, and \ceto. %
The observed clump emission traced by both \tco\ and \hcop\ (from Phase I) shows very similar morphology, indicating that, for dense molecular clouds and complexes {\em of all sizes}, parsec-scale clumps %
contain $\Xi$ $\sim$ 75\% of the mass, while only 25\% of the mass lies in extended (\gapp10\,pc) or ``low density'' components in these same areas.  The mass fraction of all gas above a density 10$^9$\,m$^{-3}$ is $\xi_9$ \gapp\ 50\%.  This suggests that parsec-scale clumps may be the basic building blocks of the molecular ISM, rather than the standard GMC concept.  Using \tco\ emission, we derive physical properties of these clumps in their entirety, and compare them to properties from \hcop, tracing their denser interiors. %
We compare the standard $X$-factor converting \ico\ to \nhtwo\ with %
alternative conversions, and show that only the latter give whole-clump properties that are physically consistent with those of their interiors.  %
We infer that the clump population is systematically closer to virial equilibrium than when considering only their interiors, with perhaps half being long-lived (10s of Myr), pressure-confined entities which only terminally engage in vigorous massive star formation, %
supporting other evidence along these lines previously published.
\end{abstract} 

\keywords{astrochemistry --- ISM: molecules --- radio lines: ISM --- stars: formation} 

\section{Introduction}

The formation of massive stars and clusters from molecular clouds remains one of the major unsolved problems in astrophysics.  Part of the reason for the ongoing debate about initial conditions, mechanisms, timescales, feedback, environmental factors, and other model parameters, is the complexity of massive star formation phenomenology and the relatively small number of wide-field, complete, uniform, unbiased, multi-wavelength studies, as explained by \citet{b11}.

For example, it has not been established whether the parsec-scale massive clumps that form star clusters \citep{ll03} are long-lived entities (several 10s of Myr) that do not undergo vigorous massive star formation until the latter part of this time span \citep[e.g.,][]{kss09,b13}, or whether they are shorter-lived objects ($<$10\,Myr) that promptly form clusters and are then dissipated \citep[e.g.,][]{bsd10}.  For long lifetimes, one must also argue for cluster-forming clumps to be either gravitationally bound or pressure-confined, as originally explained by \citet{bm92} based on stability arguments.  On the other hand, even unbound clouds could form some stars \citep{wws14}.

More recently, observational breakthroughs have provided new challenges to theory.  The {\em Herschel} Observatory revealed widespread filamentary structures in the cold interstellar medium (ISM), and subsequent theoretical work is confirming how important filaments are to molecular clouds' overall physics and star formation activity \citep[see review by][]{a14}, including how gas flows assemble filaments and then clumps within them.  Yet, many studies to date have concentrated on a few, typically nearby clouds.  Similarly, there is a growing recognition that ``CO-dark gas'' \citep{gct05,p13} may contribute up to half the molecular mass of the Milky Way.  Our direct knowledge of the distribution of this gas in the Galactic disk is currently limited to $\sim$0\fdeg2 scales \citep{L14}, yet new models suggest its distribution is related to the much finer scale of filaments \citep{s14}.  In general, many questions about molecular cloud stability, dynamics, composition, and star formation activity would best be examined with high spatial dynamic range imaging of a wide sample of Giant Molecular Clouds (GMCs).

We developed the Galactic {\em Census of High- and Medium-mass Protostars} (CHaMP) to address many of these issues.  CHaMP was originally conceived to be a multi-wavelength survey of a statistically large but uniformly-selected sample of massive, dense molecular clumps, in order to analyse the massive star- and cluster-formation process with as little observational bias as possible.  CHaMP began with a Mopra\footnote{The Mopra telescope is part of the Australia Telescope, funded by the Commonwealth of Australia for operation as a National Facility managed by CSIRO. The University of New South Wales Digital Filter Bank used for observations with the Mopra telescope was provided with support from the Australian Research Council.} 
molecular line mapping survey of a complete, flux-limited clump sample within a 20\degree$\times$6\degree\ area of the southern Milky Way in Vela, Carina, and Centaurus.  Phase I of this mapping, of the emission from several ``dense\footnote{What exactly constitutes a dense clump is not always well-defined in the literature; we explore this in \S\ref{clumpcomps}.} 
gas tracers'' in the identified CHaMP clumps, took place over the period 2004--2007, with several follow-up studies at infrared wavelengths since then, using the AAT, CTIO, {\em Spitzer}, and archival spacecraft data.  Early results from this work have been presented by \cite{y05}, \cite{z10}, \citet{b10}, and \citet{b13}, with more studies in preparation.

In \citet[hereafter Paper I]{b11}, we described the overall survey strategy and reported the first mm-wave results for the ensemble of 303 massive molecular clumps in the southern Milky Way.  We found that these clumps represent a vast population of subthermally-excited, yet massive clouds, 95\% of which are relatively quiescent and not currently engaged in vigorous massive star formation, suggesting long clump lifetimes.  In \citet[hereafter Paper II]{mtb13}, we performed a global analysis of the spectral energy distributions (SEDs) of these clumps, and found a wide range of star-formation efficiencies (SFEs), consistent with such long lifetimes.

At the same time, it was recognised that similarly-detailed information on the clump envelopes and embedding GMCs would be needed to provide critical comparisons with the denser gas, such as abundances, masses, and an environmental context for any star formation activity within the clumps.  Thus, Phase II observations, primarily aimed at mapping the CO-isotopologue lines with Mopra, were conducted during 2009--2012.  Because of its uniform and wide-area approach, CHaMP's strategy of fundamental cloud demographics was designed to access new discovery space in the pursuit of the science problems described above.

In this sense, CHaMP differs strongly from several other molecular ISM/star formation projects.  For example, detailed studies on individual massive star formation sites \citep[e.g.,][for OMC1, NGC\,6334, and DR21, resp.]{u97,r10,s10} find molecular gas properties which are more extreme than any members of the CHaMP sample.  Surveys of massive star formation may also be selective: e.g., both \citet{wu10} and the MALT90 project \citep{j13,h13} examined more extreme samples of clouds than CHaMP, as discussed in Paper I.  The former targeted $\sim$50 of the most luminous water masers in the Galaxy, while the latter mapped $\sim$2000 of the highest column density, pc-scale dust clumps in the southern Galactic Plane from the 870\,$\mu$m ATLASGAL \citep{s09} clump catalogue.  In both cases, such projects specifically pick out the most extreme cloud population (e.g., in terms of luminosity, opacity, mass, or density), rather than the much more representative, and ultimately much larger and less biased, CHaMP cloud population, which probably numbers $>$10$^4$ across the Galaxy.  Therefore, it should not be surprising that the cloud properties we find are different than in these other samples, even when observed with the same telescope.

In this paper we describe the Phase II observing (\S2) and data reduction (\S3) procedures, paying particular attention to where these differ from Phase I, and give the first results from analysing the brightest Phase II line, \tco\ \joz, in \S4.  We discuss these results in \S5, in the context of our previous results on the dense gas, IR emission, and star formation activity in these clumps, while also relating this to the current wider understanding of the cluster formation process.  Our conclusions are summarised in \S6.

\section{Observations}\label{observe}

In Phase II of the Mopra observing for the CHaMP project, we tuned the receiver to a central frequency of 111.3\,GHz and set up the MOPS digital filterbank %
to map {\bf{\em all the CHaMP clumps}} in a second set of spectral lines at frequencies 107--115\,GHz.  This new set of transitions most notably includes the \joz\ lines for the triad of CO-isotopologue species, \tco, \ttco, and \ceto.  In contrast, Phase I of the observing featured transitions from a number of bright dense gas tracers near 90\,GHz, and was described in Paper I with results and analysis of the brightest of these lines, \hcop.  Additionally, we presented results on a subset of the \nnh\ line data in \citet{b13}, also from Phase I. %


Apart from observing a different set of spectral lines, the observational procedures in Phase II were similar to those used during Phase I, and the interested reader is referred to Paper I for those details.  The Phase II mapping was designed to completely cover the spatial extent of the same ``Regions'' (ranging in size from $\sim$0\fdeg1--1$^\circ$; see Paper I or Appendix A for definition) as were mapped in Phase I, but we expected that, for the brighter \tco\ and \ttco\ lines, the molecular cloud emission was likely to occur over a somewhat wider area than was seen for even the brightest of the Phase I lines (\hcop).  Thus, the Phase II maps are generally somewhat larger than the equivalent areas shown in Paper I, although due to the usual observing constraints of limited time and adverse weather, this ideal was not achieved in all cases.

Especially for the \tco\ and \ttco\ lines, we also understood that the selection of emission-free positions for sky subtraction in Mopra's on-the-fly (OTF) mapping would be important to enable high-quality analysis of the map data.  We used the existing lower-resolution but wider-scale Nanten maps (see Paper I) to identify positions where the \tco\ and \ttco\ emission was undetectable down to the Nanten noise limit, but in about half the Regions, the OFF position selected was less than ideal.  Sometimes this was recognised during the early mapping for each Region, a better OFF position selected, and the mapping redone.  In most cases this worked well, in the sense that more than half the resulting Region maps have no detectable features from emission in the reference beam (ERB).  In the other maps, ERB is still present, but it is almost always quite weak $\sim$0.1--0.7\,K), and usually at velocities that are quite different from those of the clouds under study.  In a few cases (Regions 8, 16, 26a, 26b), these precautions were not as effective, and in \S3 we describe the analysis procedures we used to mitigate these effects.

\begin{figure}[t]
\includegraphics[angle=0,scale=0.45]{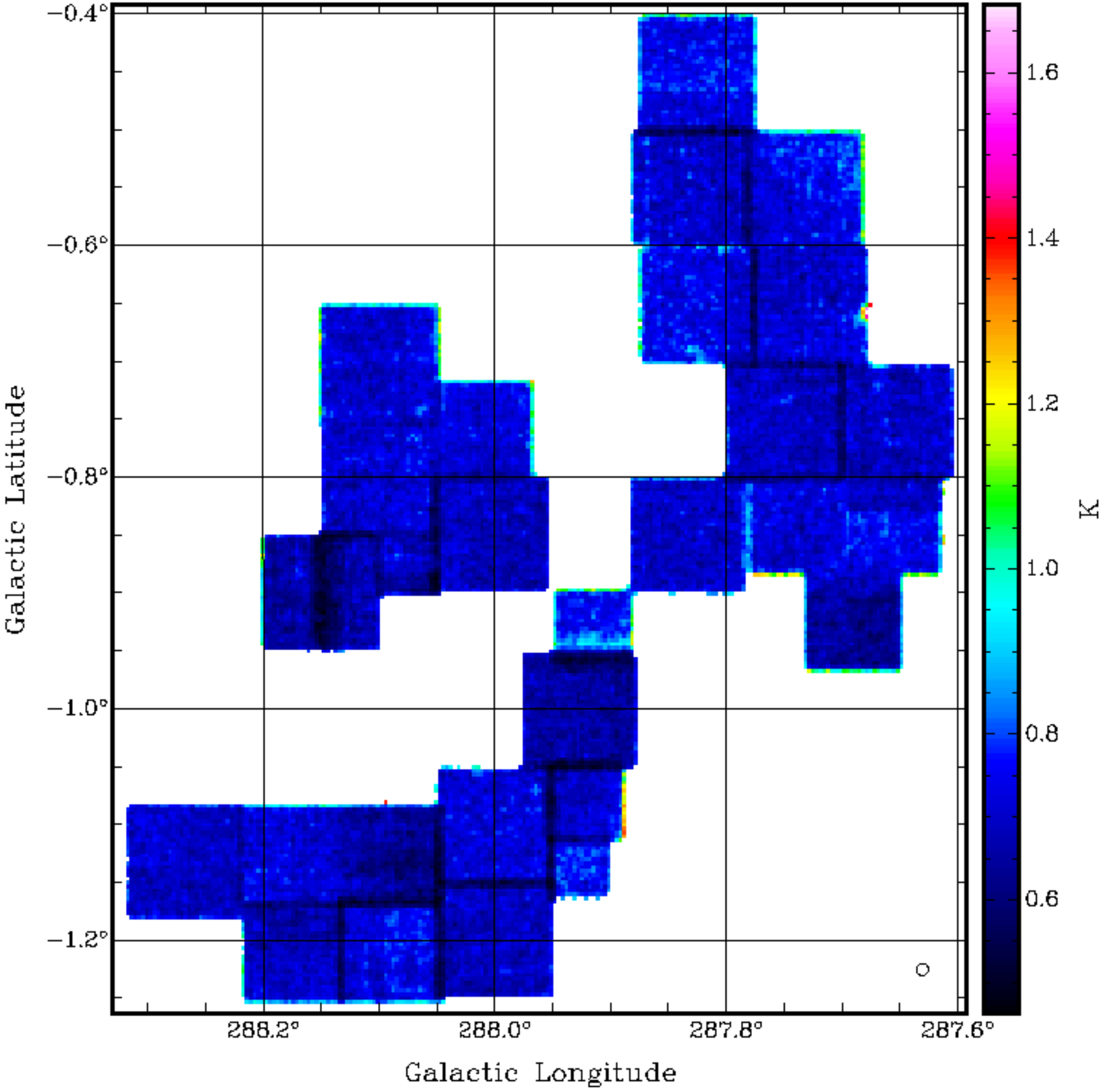}
\caption{Sample map of rms/channel 
at each pixel for Region 11.  The Mopra beam is shown in the BR corner.  Although the boundaries between separately mapped fields are clearly visible as narrow, lower-noise bands where the fields overlap, the overall noise distribution across all fields is quite narrow, in this case 
0.7$\pm$0.1\,K.
}
\label{rmsmap}
\end{figure}

In Phase II, we also used a new technique which we call ``Active Mapping'' (AM) to compensate for slow ($\sim$hour-long) variations in atmospheric conditions.  This technique, described in detail by \citet{bd09} and \citet{bm15}, uses a ``Nyquist frequency'' \nf\ to set the OTF sampling speed on the sky in such a way as to greatly reduce most of the noise variations in large maps with multiple fields, due to the inevitably different observing conditions on different days, or due to variations in \tsys\ due to elevation (i.e., airmass).  In short, if the \tsys\ rises, then \nf\ is scaled higher in proportion; this proportionately slows down the OTF scan speed in {\bf{\em both}} dimensions, and gives an integration time per pixel that varies as \tsys$^2$, effectively cancelling out the higher noise from the higher \tsys:
\begin{equation}   
	{\rm rms}~{\rm noise} \propto \frac{T_{\rm sys}}{\sqrt{t_{\rm int}}} \propto \frac{T_{\rm sys}}{\sqrt{\nu_N^2}} \propto \frac{T_{\rm sys}}{\sqrt{T_{\rm sys}^2}} \propto {\rm constant} .
\end{equation}
The resulting noise distribution in maps made with AM is much narrower than in those made with standard OTF mapping \citep[see][and Fig.\,\ref{rmsmap}]{bm15}.

As with Phase I, pointing was checked during the observations every hour or so, using the SiO maser R Carinae \citep{i13}.  %
Also similarly to Phase I, the Mopra system was found to be quite stable overall, although there were variations in the seasonal calibration averages by $\pm$15\% \citep[see][their Figure 5, for an example from the ThrUMMS project, which was observed over the same time period]{bm15}.  These factors were measured and used to put the data from different seasons onto a common brightness scale (see \S\ref{dataproc}).  It should also be noted that, over the 8\,GHz frequency range of the line data, the atmospheric opacity changes strongly from 107\,GHz to 115\,GHz, such that (e.g.) the \ceto\ and \ttco\ data have noise figures which are always roughly {\em half} that for the \tco\ data over the same mapped areas.

\notetoeditor{}

\section{Data Reduction and Processing}\label{dataproc}

As in Phase I, we use the \textsc{Livedata-Gridzilla} package \citep{b01} to perform the standard Mopra data reduction.  However, for this release we have enhanced the standard pipeline in a number of ways, improving both the final quality in the reduced data cubes, as well as the fidelity of the analysis products derived from the cubes.  We refer the reader to Paper I for the standard treatment, while the improvements have been described by \citet{bm15}.  We nevertheless give a brief summary here.

First, \textsc{Livedata} was used as before to extract the \tco\ %
spectral line data %
from the raw data file, perform a preliminary calibration and baselining, and write the result to a normal single-dish FITS file.  In this processing, a fraction of the data ($\sim$2\% overall, but ranging from 0--20\% in any given file) were found to be discrepant in various ways (anomalous calibration, bad baseline division, etc.).  Left uncorrected, such data can produce noticeable deleterious effects in the maps, such as edges, ``bright'' or ``dark'' spots, warped baselines, and so on.

Many of these problems are easy to miss with a na\"ive application of the standard \textsc{Livedata-Gridzilla} pipeline.  In such cases, the deleterious effects of the bad data on the maps persist, even when they aren't obvious, e.g., where they are masked by the coaddition of ``good'' data.  Such problems are apparently due to temporary hardware or software malfunctions during the observations or data reduction, and were first described by \citet{bm15} as part of ThrUMMS, but their root cause has not been identified.  Nevertheless, once one is aware that these effects can occur, they are easy to uniformly screen for in the raw data, identify, and then either flag or correct. %

We developed custom software to perform this remediation, written as a combination of IDL and Unix c-shell scripts, and effected between the \textsc{Livedata} and \textsc{Gridzilla} stages of the processing \citep{bm15}.  This is also the stage at which we applied the seasonal calibration factors described in \S\ref{observe}.  These additional calibration and editing steps improved the overall data quality to the point where the resulting noise levels were always consistent with theoretical expectations, given the observing conditions.

\notetoeditor{}
\begin{figure*}[ht]
\hspace{88mm}\includegraphics[angle=0,scale=0.33]{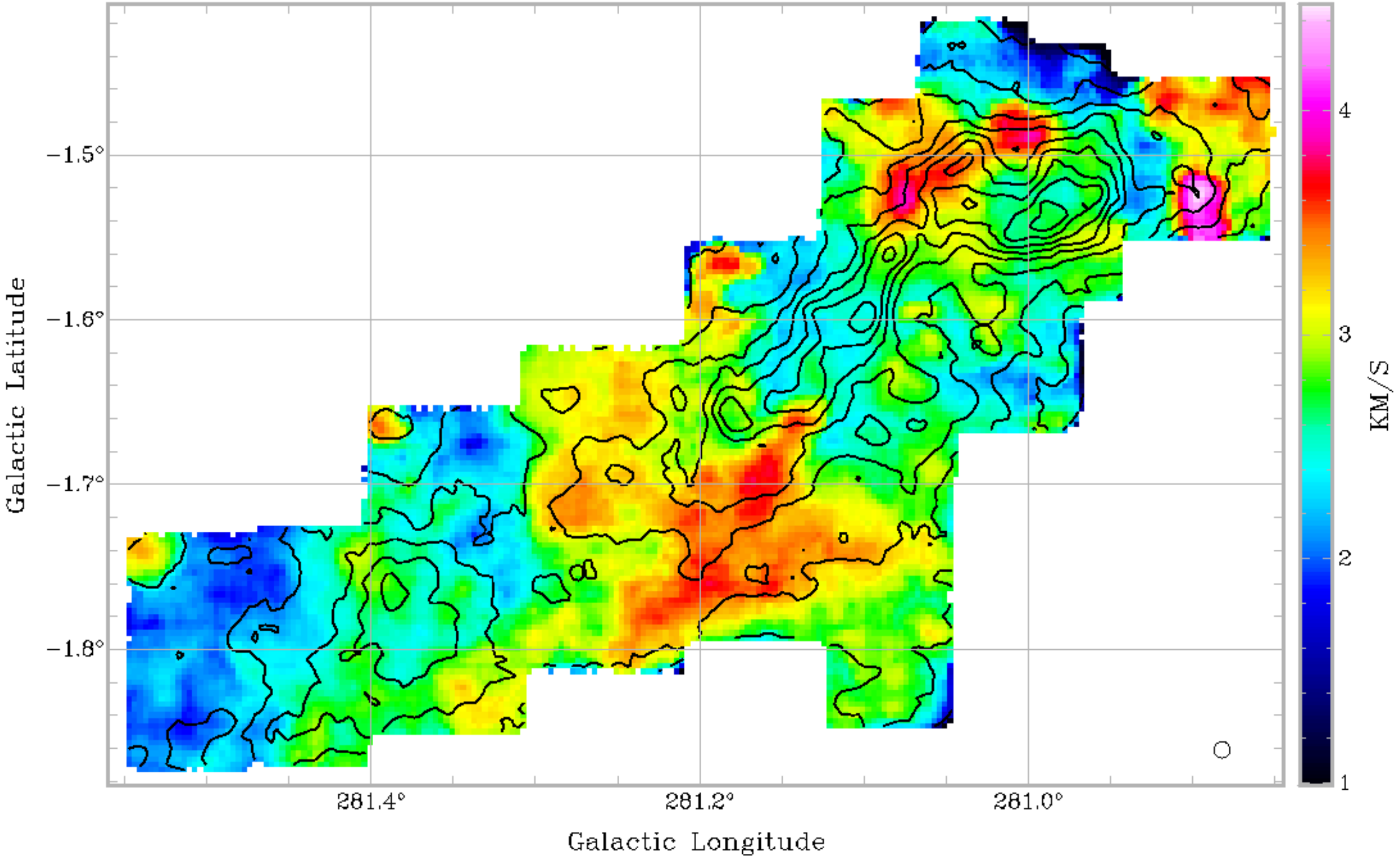}
\hspace{-181mm}\includegraphics[angle=0,scale=0.33]{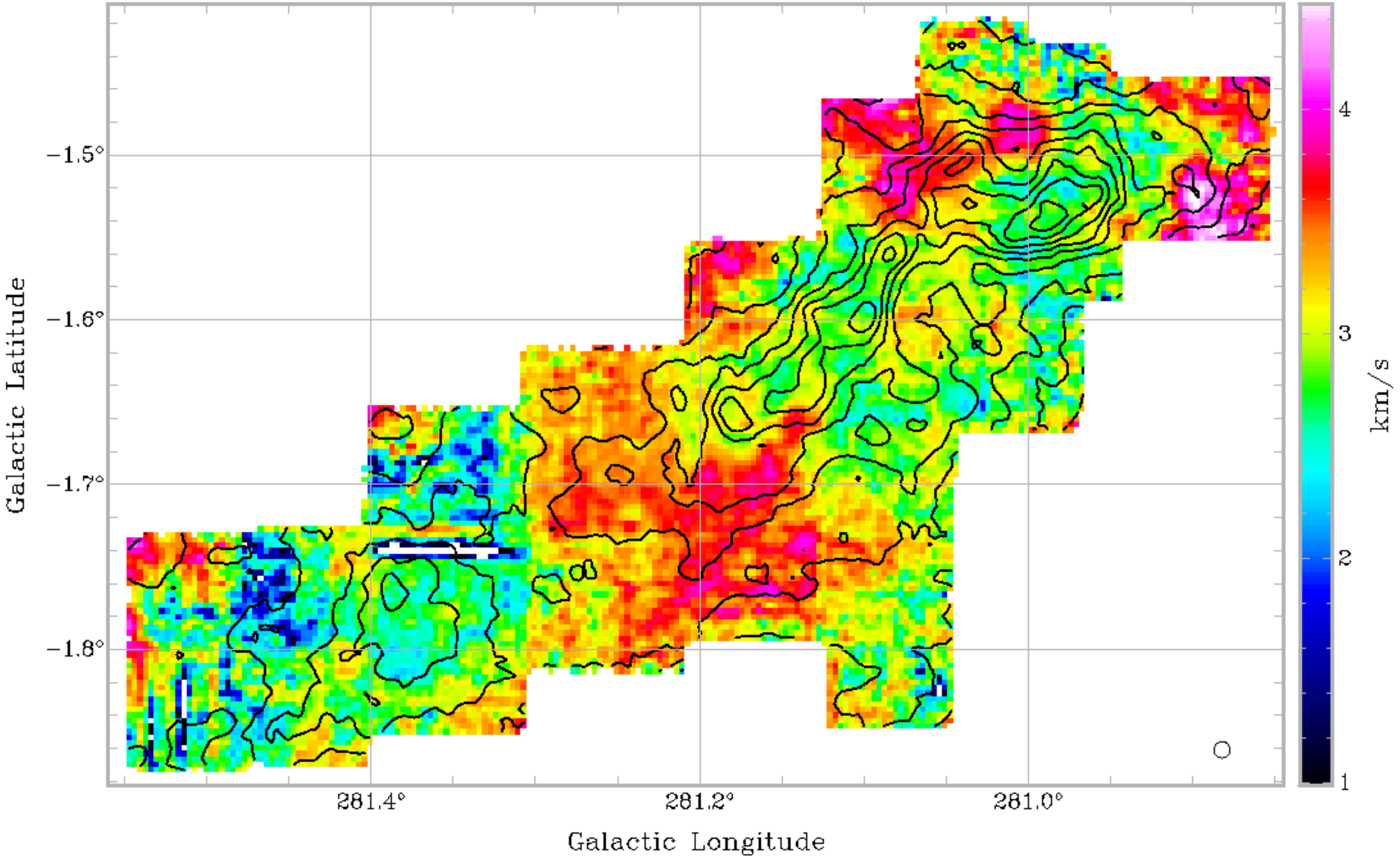}
\caption{Sample velocity dispersion (2nd moment) maps for part of Region 1.  Both maps are overlaid by the same contours of integrated intensity (0th moment) from the SAMed data.  The Mopra beam is shown in the BR corner.  (left) Computed with all data inside a velocity range of --15.5 to +3.0\,\kms, and with uncorrected bad baselines and mis-scaled data %
(see \S\ref{dataproc}) in some locations.  (right) Computed over the same velocity range as the left panel, but using our SAM technique \citep{bm15} to omit low-significance voxels from the moment calculation, and with corrections for known data errors.  Both panels are placed on the same brightness scale to facilitate comparisons.  Note the dramatically reduced impact of beam-sized noise features in the right-hand map, and the consequent improvement in self-consistency, fidelity, and image quality, even for the \tco\ line which has high S/N in most locations.  The SAM technique makes these improvements across all areas of the map, whereas the data flagging/error correction makes much larger improvements, but only over smaller areas where the errors occur (e.g., the striations in the BL portion of the map).  %
Improvements with SAM are even more noteworthy for weaker lines.
}
\label{mom2sample}
\end{figure*}

\textsc{Gridzilla} was then used to produce data cubes covering the full extent of the region mapped, and over a standardised \vlsr\ range of --60\,\kms\ to +40\,\kms\ (although each 138\,MHz wide IF zoom actually covers about 360\,\kms; based on the Nanten data, only the narrower velocity range contains detectable emission).   The maps have an effective HPBW of 37$''$ for the \tco\ data reported here, slightly smoothed from the intrinsic 33$''$ of the telescope at this frequency \citep{L05}.

From the data cubes, moment maps were then calculated; these are a convenient tool for extracting properties of the emission regions, as was used in Paper I.  Here we developed an improved pipeline to calculate these moment maps, one which was built around the {\em smooth-and-mask} (SAM) method for the ThrUMMS project \citep{bm15}, but optimised here for the CHaMP data.  SAM techniques have a dramatic effect on improving the quality of moment maps, especially at low S/N, compared to those made with simple velocity-limited integrations, since they are designed to automatically include data only where they are significant, regardless of preconceived definitions of source area or velocity extent.  %
An example of SAMing is given in Figure %
\ref{mom2sample}, where one can see a substantial improvement in data quality compared to Paper I (i.e., far better than expected from just the higher S/N in the brighter \tco\ data compared to \hcop); see \citet{bm15} for more details.

In the few Region maps (8, 16, 26a \& b) where ERB occurs at velocities that are close to the cloud emission of interest, we needed to compensate for its presence before the measurement of parameters as described in \S\ref{results}.  We did this by forming an average spectrum over an area of the mapped Region with minimal or no emission from the cloud of interest.  This average ERB spectrum was consequently of high S/N, even for the weak ERB feature(s).  We then fit the ERB with a simple gaussian line of negative amplitude (one component was always enough), and added this component back into the data cube.  In this way, we are confident that any residual effects from ERB in our maps are close to the maps' respective noise levels.

Finally, we can use the moment maps to check the overall calibration of the Mopra data and the image fidelity in our maps.  We tried to compare our \tco\ maps with those from the Columbia-CfA CO survey \citep{dht01}, as was done by \citet[][their Figure 6]{bm15} for the ThrUMMS project.  However, at Mopra's full angular resolution of 0\farcm6, and with much smaller CHaMP map sizes than for ThrUMMS, the comparison to the much (14$\times$) lower resolution CfA survey becomes very sparse, since the effective pixel size in suitably convolved CHaMP maps needs to be 186$\times$ larger.  Instead, we make this comparison in two steps, using the Nanten maps (which have resolution 3\farcm3, only 5.4$\times$ lower than Mopra; see Paper I) as an intermediary. %

With this approach, and allowing for the sub-Nyquist sampling in the Nanten maps, we obtain a relative calibration of (CHaMP)/(CfA) = 0.975$\pm$0.042, based on an arbitrary subset of all our \tco\ maps.  This excellent result shows that our calibration techniques are very reliable, not just for \tco, but (because of the simultaneity of the multi-species map-making afforded by the MOPS backend) for {\bf{\em all}} the Phase II spectral line maps made at Mopra.  However, we emphasise that, without the extra steps described in this section and \S\ref{observe}, the calibration would have been much less reliable.

In Appendix A, we present all the \tco\ moment maps over all observed Regions and isolated clumps listed in the Nanten Master Catalogue of Paper I.

\section{Results}\label{results}

\subsection{Preamble on the $X$-factor}\label{xfactor}

A recurring theme in the remainder of this paper is the issue of the conversion of \tco\ line intensities to column densities, via the $X$-factor or some other method.  The $X$-factor is calibrated observationally via %
various estimates of total HI+\htwo\ column density; correlations were also found between virial masses and CO luminosities, yielding similar $X$-factors \citep[e.g.,][and references therein]{dht01,bwl13}.  Yet, historically the $X$-factor has not been well-calibrated in high column density settings \citep{bm15}.  Thus, the view arose that the $X$-factor works, in the mean, because the mass is ``encoded'' by the \tco\ linewidth \citep{bwl13}, despite the high line opacity and the disparate values derived in different studies, with cloud masses both well above and well below virialised levels \citep[][Paper I]{bm92,n09,rd10}.

We argue instead that such a view is an oversimplification, and that the ThrUMMS law not only explains the origin of the $X$-factor in terms of fundamental radiative transfer physics (it depends on 3 factors, the \tco\ line's $\tau$, $\sigma$, and \tex), but also provides a more robust and consistent conversion in all settings, allowing for clouds' excitation (environment), and independent of their virial state.  In what follows, we explore this issue in four steps:\\
\hspace*{3mm}1. In \S\ref{physpar}, for demonstration purposes we derive clump parameters based on the measured cloud properties and a standard $X$-factor.\\
\hspace*{3mm}2. In \S\ref{coplots}, we find the clumps are not in virial equilibrium, but the properties are subtly inconsistent with the \hcop\ results from Paper I.\\
\hspace*{3mm}3. In \S\ref{newx}, we derive clump properties with the same observed parameters but use alternative and independently calibrated conversion laws instead of the usual $X$.\\
\hspace*{3mm}4. In \S\ref{clumpcomps}, we find that the clump properties are more consistent with the Paper I \hcop\ results, {\em and} that when accounting for the envelope mass this way, the clumps are much closer to virial equilibrium than seen in step 2.

\subsection{Observed Clump Parameters in \tco\ }\label{obspar}

\begin{figure}[ht]
\includegraphics[angle=0,scale=0.46]{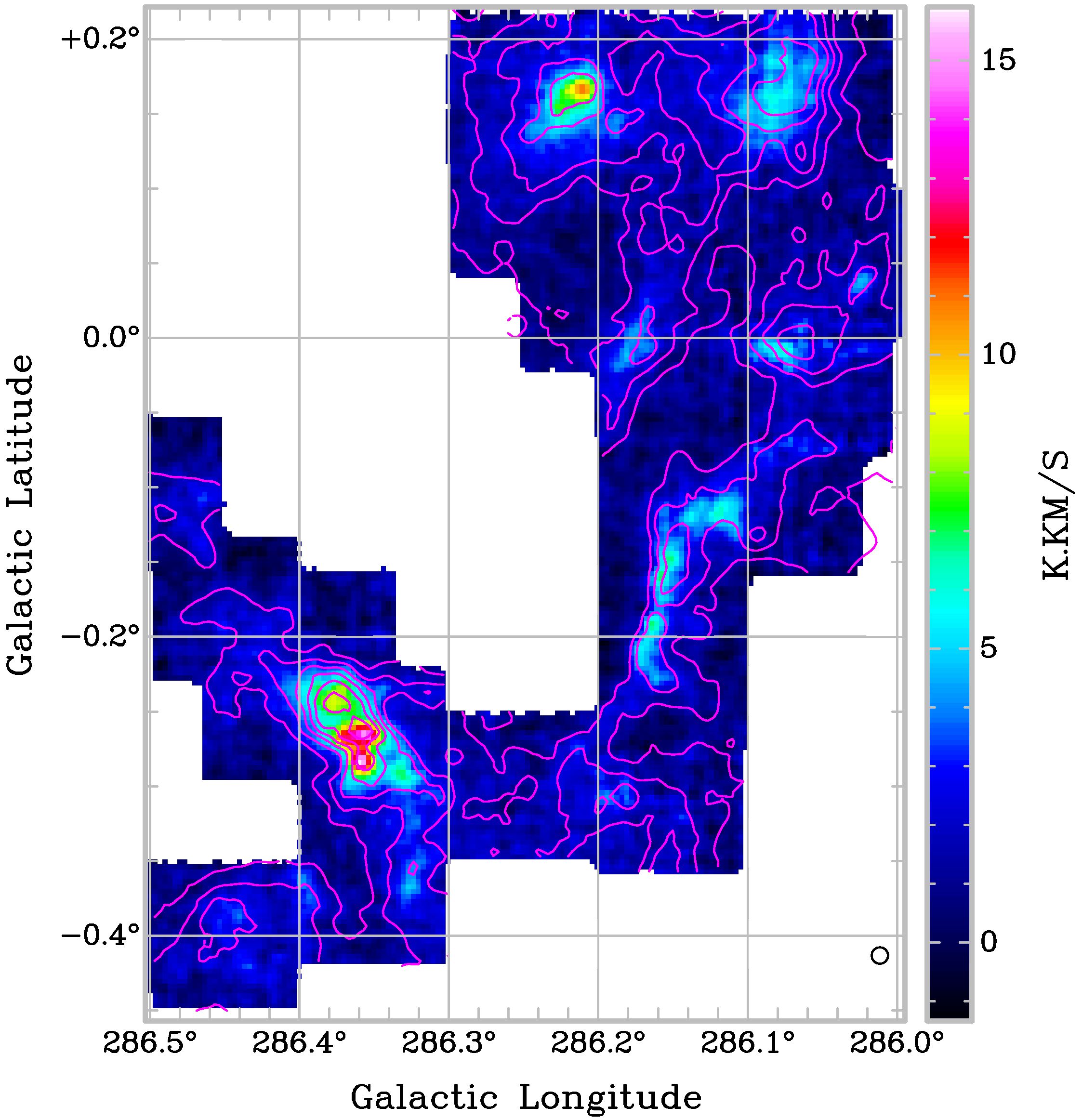}
\caption{Overlay of \tco\ contours (magenta, every 38$\sigma$ = 21.3\,K\kms) on an \hcop\ image of part of Region 9.  Note the close correspondence between the emission morphology in the two species.  See text for discussion.
}
\label{overlay}
\end{figure}

Rather than measure the clump parameters based solely on the high optical depth (see \S\ref{physpar}) \tco\ maps, we use the measured positions, sizes, and orientations of the \hcop\ clumps from Paper I to define ``areas of interest'' in these \tco\ maps, and derive comparable observed quantities ($T_{\rm peak}$, $I_{\rm CO}$, \vlsr, $\sigma_V$, sizes, shapes) for the 303 identified clumps of Paper I.  The rationale for this is that the \hcop\ is nearly always optically thin,\footnote{This is true in the CHaMP sample (see Paper I), but unlike the clouds in the MALT90 sample, which have systematically higher opacities \citep{h13}.} 
and so should be more representative of the total gas column in each cloud, as opposed to the clouds' expected ``surface'' properties, likely to be traced by the very optically thick \tco.  In most cases however, the emission morphology between the \tco\ and \hcop\ is very similar, so this distinction is mostly  moot.  %
Figure \ref{overlay} gives an example of how this similarity pervades nearly all of the mapped structures in these two molecular lines; this relationship is quantified in \S\ref{obscomps}.

More importantly, we are interested here in the properties of the molecular clump envelopes that contain the denser gas traced by the \hcop\ emission, rather than {\bf{\em all}} the detectable \tco\ emission.  Therefore, we confine the discussion below to the \tco\ properties of {\bf{\em only}} those \hcop\ clumps previously catalogued in Paper I, and analysis of the several hundred other features visible in the \tco\ maps, but not detected in \hcop, is deferred to a future study.

\notetoeditor{}

The \tco\ \joz\ data for these CHaMP clumps are presented in Table B1 in a similar format to the \hcop\ results in Table 4 of Paper I, but with the addition of 2 extra columns.  Column 1 gives the clump designation from Paper I, with columns 2--3 giving the position of the \tco\ $I_{\rm CO}$ emission peak nearest to the corresponding \hcop\ coordinates, and the $I_{\rm CO}$ peak value in column 4.  Column 5 shows the velocity range used for the moment calculations, with the main-beam brightness temperature, intensity-weighted mean velocity, and velocity dispersion, all measured at the peak $I_{\rm CO}$ position, given in columns 6--8.  Using the procedure described in Paper I for decomposing the emission into 2D elliptical gaussian components, shape parameters approximating the \tco\ emission morphology are given in columns 9--13, including the emission centroid (columns 12--13) based on the half-power contour (columns 9--10), as opposed to the emission peak in columns 2--3.  These last 2 columns are additional to those in Paper I.  %

Although the \tco\ and \hcop\ morphologies do correspond closely to each other for most clumps, there are a few situations where they don't match very well.  Sometimes, this is where two \hcop\ clumps form a single blended structure in \tco.  Such cases are shown in Table B1 with dual subclump designations, such as ``7ab''.  The number of such blended structures is quite small compared to the total (10/273, or 3\% of the total), again underscoring the overall structural similarity.

\subsection{Derived Physical Parameters of \tco\ Clumps}\label{physpar}

The \tco\ molecule is very abundant ($\sim$10$^{-4}$ compared to \htwo, see below), and emission from the easily-excited \joz\ line is certain to be very opaque ($\tau$$\gg$1, probably several 10s or more) almost everywhere it is seen.  This means that maps of \tco\ emission are essentially tracing the molecule's excitation temperature \tex\ at the surface of last scattering on the front side of each molecular cloud, or to be more exact,
\begin{eqnarray}   
	T_{\rm mb} & = & (T_{\rm ex}-T_{\rm bg})(1-e^{-\tau}) \\ 
			& = &  (T_{\rm ex}-T_{\rm bg})\hspace{12mm}{\rm where}~\tau \gg 1  ,
\end{eqnarray}
where we use the Rayleigh-Jeans approximation at 3\,mm, $J\propto T$.

\notetoeditor{}
\begin{figure*}[ht]
(a)\hspace{-4mm}\includegraphics[angle=0,scale=0.45]{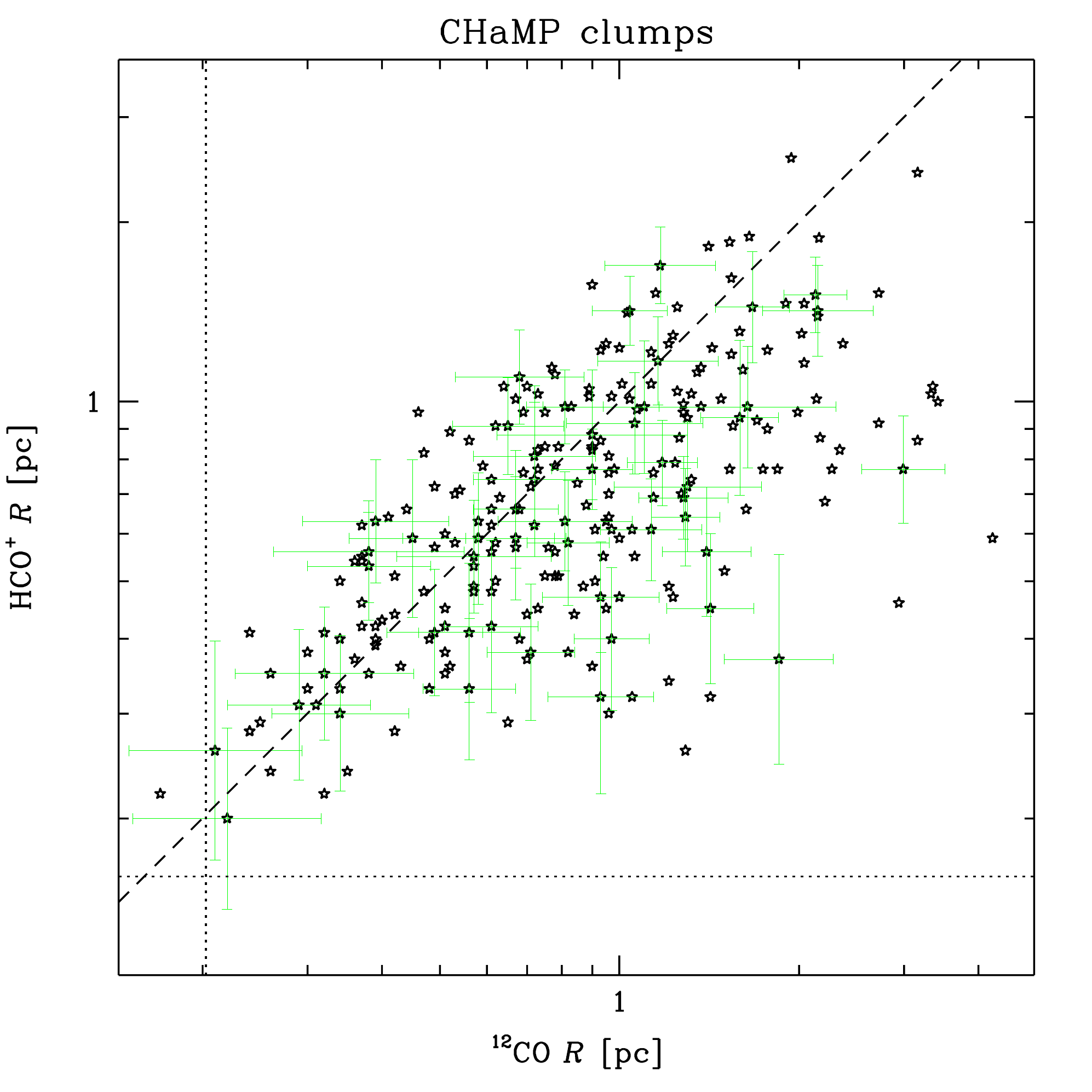}\hspace{2mm}
(b)\hspace{-4mm}\includegraphics[angle=0,scale=0.45]{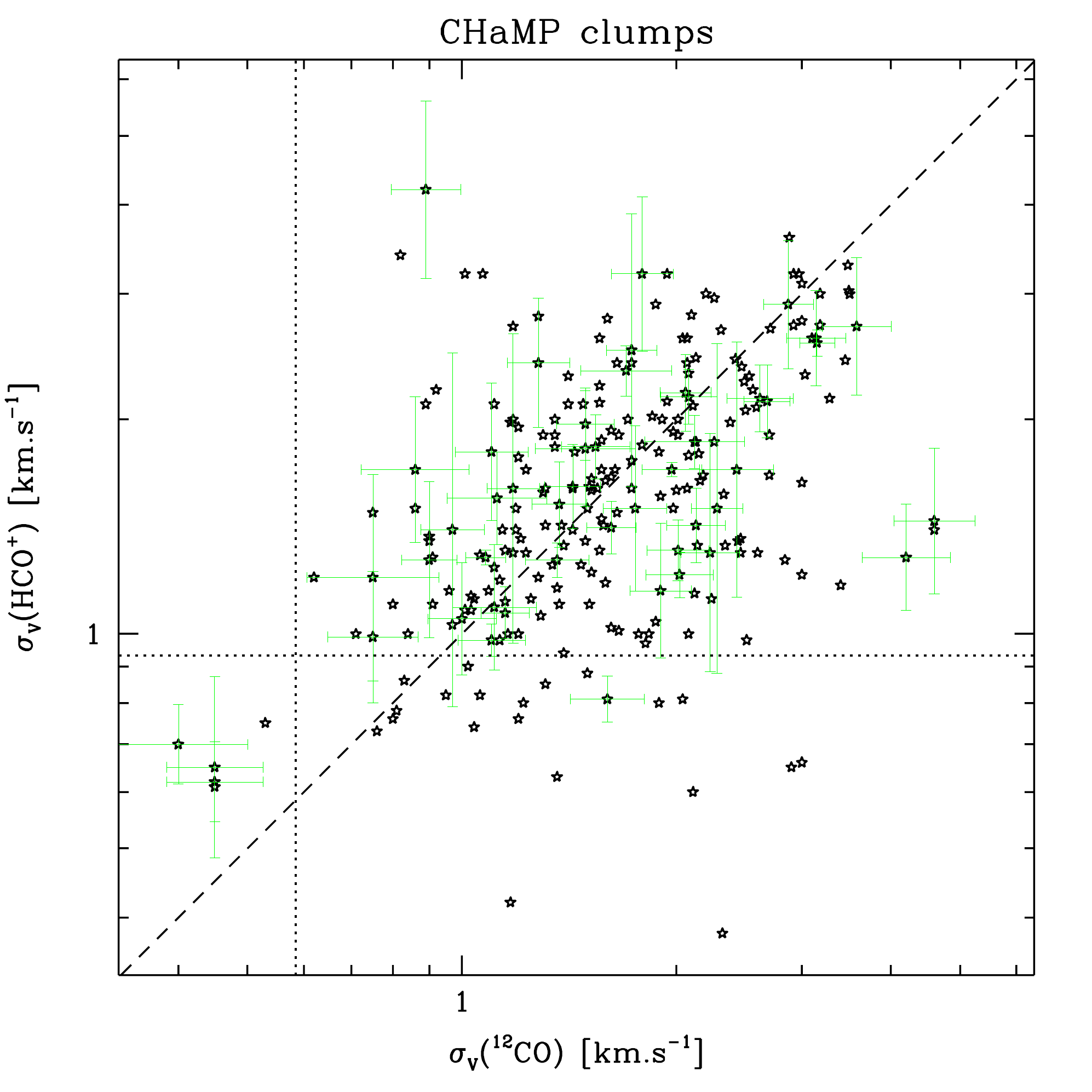}
\caption{Comparison of CHaMP clumps' size (a) and velocity dispersion (b) derived from \hcop\ measurements in Paper I ($y$-axis in both panels) and that derived from \tco\ measurements presented here ($x$-axis in both panels).  Both panels show green error bars for uncertainties on 1-in-5 points, the mean 3$\sigma$ sensitivity limits (resolution in this case) as dotted lines, and a diagonal dashed line where both species would give the same values for each parameter.  Weighted least-squares fits (not shown) are (a) log\,$R_{\rm HCO^+}$ = (0.66$\pm$0.03) log\,$R_{\rm ^{12}CO}$ -- (0.155$\pm$0.017) with correlation coefficient $r^2$ = 0.65, and (b) log\,$\sigma_V$(\hcop) = (0.62$\pm$0.04) log\,$\sigma_V$(\tco) + (0.067$\pm$0.015) with correlation coefficient $r^2$ = 0.43.
}
\label{RDVcomps}
\end{figure*}

The treatment of Paper I depends on the fundamental provision of a known (or fixed) excitation temperature \tex, and the use of this to calculate an estimate for the optical depth $\tau$ of the line under study.  This is acceptable for \hcop\ since Paper I showed that derived physical quantities are only weakly dependent on the value assumed for \tex\ if the integrated intensity $I_{\rm HCO^+}$ is known.  Also, the \hcop\ emission is never too bright to prevent calculation of a reasonable $\tau$ by inversion of eq.\,(2), since \tmb(\hcop) $<$ (\tex--\tbg).  %

In the case of \tco, we can't use this procedure: the large optical depth is difficult to estimate, since the \tco\ line brightness satisfies eq.\,(3) almost everywhere.  Without a way to determine the optical depth or column density in the \tco\ line (such as \ttco\ and \ceto\ line ratio information; Barnes et al, in prep.), further analysis of the physical conditions in the clump envelopes traced by \tco\ would necessarily be limited.

Instead, \ico\ is widely used in the literature as a measure of the \htwo\ column, based on the \xco\ factor which has been calibrated in many studies \citep[e.g.,][]{dht01}.  We therefore use the same conversion here to estimate the molecular column density\vspace{0mm}
\begin{eqnarray}
	N_{H_2} & = & X_{\rm CO}~I_{\rm CO} \nonumber \\ 
		& = & 1.8\times10^{24}~{\rm H}_2\,{\rm mol\,m}^{-2}~(I_{\rm CO}/{\rm K\,km\,s}^{-1})
\end{eqnarray} 
or the mass surface density\vspace{0mm}
\begin{eqnarray}   
	\Sigma_{\rm mol} & = & N_{H_2}~\mu_{\rm mol}~m_{\rm H} \nonumber \\ 
	& = & 3.38\,{\rm M}_{\odot}{\rm pc}^{-2}~(I_{\rm CO}/{\rm K\,km\,s}^{-1}) 
\end{eqnarray}
in the CHaMP \tco\ maps, where $\mu_{\rm mol}$ = 2.35 is the mean molecular weight for an atomic He abundance of 9\% by number, and \ico\ is the observed intensity in each voxel of velocity width d$v$. 

Eq.\,(4) represents an average conversion factor measured over a wide range of cloud conditions; the actual \tco\ gas-phase abundance with respect to \htwo\ is likely to be variable.  For example, a variety of observational measurements and chemical models \citep[see][and references therein]{gbs14} indicate that the fractional \tco\ abundance relative to \htwo\ typically lies in the range 10$^{-3.5}$ to 10$^{-4.5}$.  The value of 10$^{-4}$ often used in the literature is only a mean over this range, but in that case, one can also compute \nco\ = 10$^{-4}$\nhtwo\ from eq.\,(4).

This factor of $\sim$3 uncertainty in the gas-phase CO abundance is typical in molecular cloud mass determinations, and is somewhat irreducible without (e.g.) detailed radiative transfer studies that are beyond the scope of the present work.  We therefore compute \nco\ and $\Sigma_{\rm mol}$, as described above, at the peak position of each clump listed in Table B1, with this uncertainty being understood.  Then, using the Paper I distances and assuming a simple 3-dimensional gaussian clump model as in Paper I, we derive each clump's radius, peak number- and mass-density, total mass based on column density, implied central gas pressure, line luminosity, virial $\alpha$ also based on column density, %
Jeans radius, and Bonnor-Ebert mass in the same manner as was done for the \hcop.  (The interested reader should refer to Paper I for details on the computational procedure.)  These results are provided in Table B2, again in similar fashion to Paper I, Table 5.

\notetoeditor{}
\begin{figure*}[ht]
(a)\hspace{-4mm}\includegraphics[angle=0,scale=0.45]{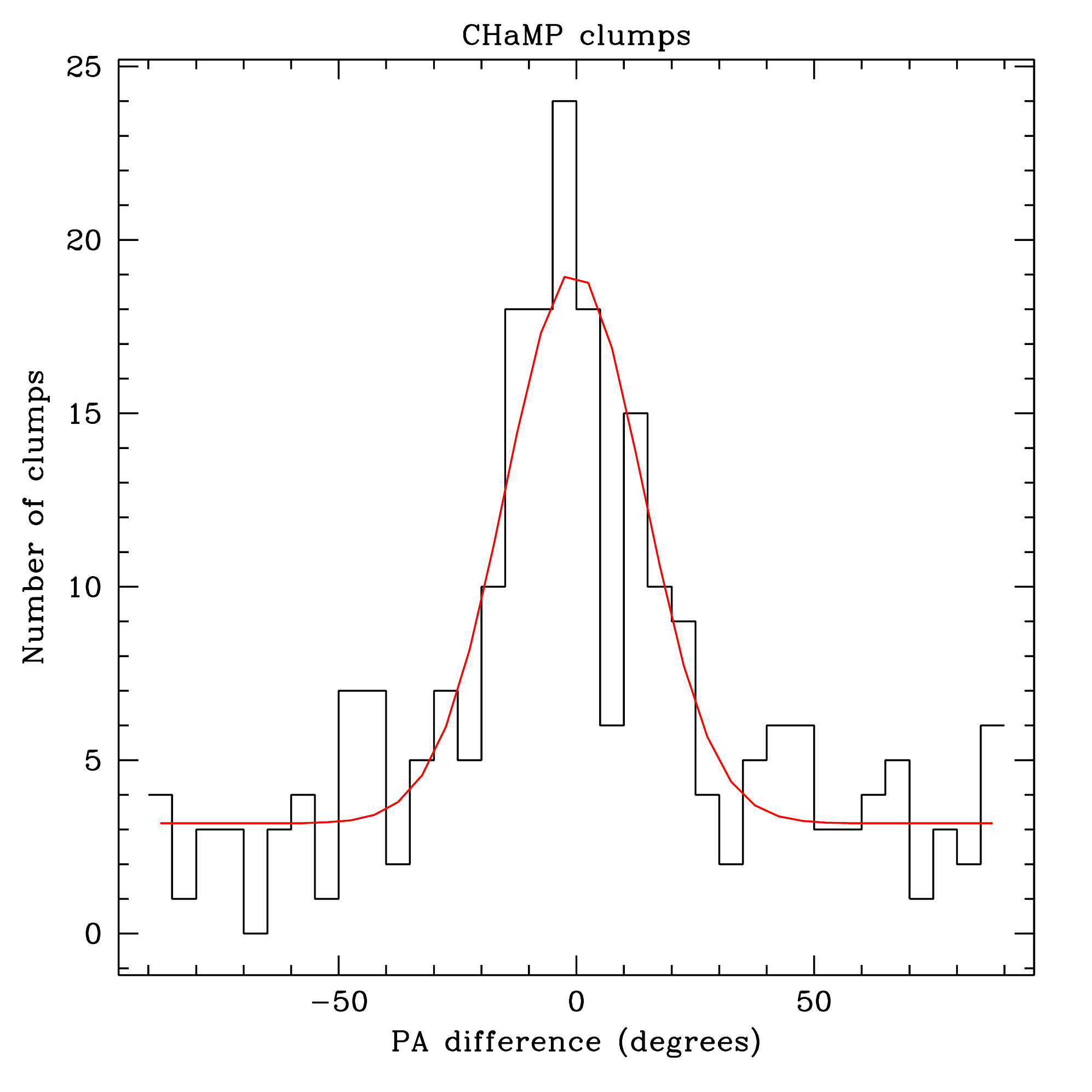}\hspace{2mm}
(b)\hspace{-4mm}\includegraphics[angle=0,scale=0.45]{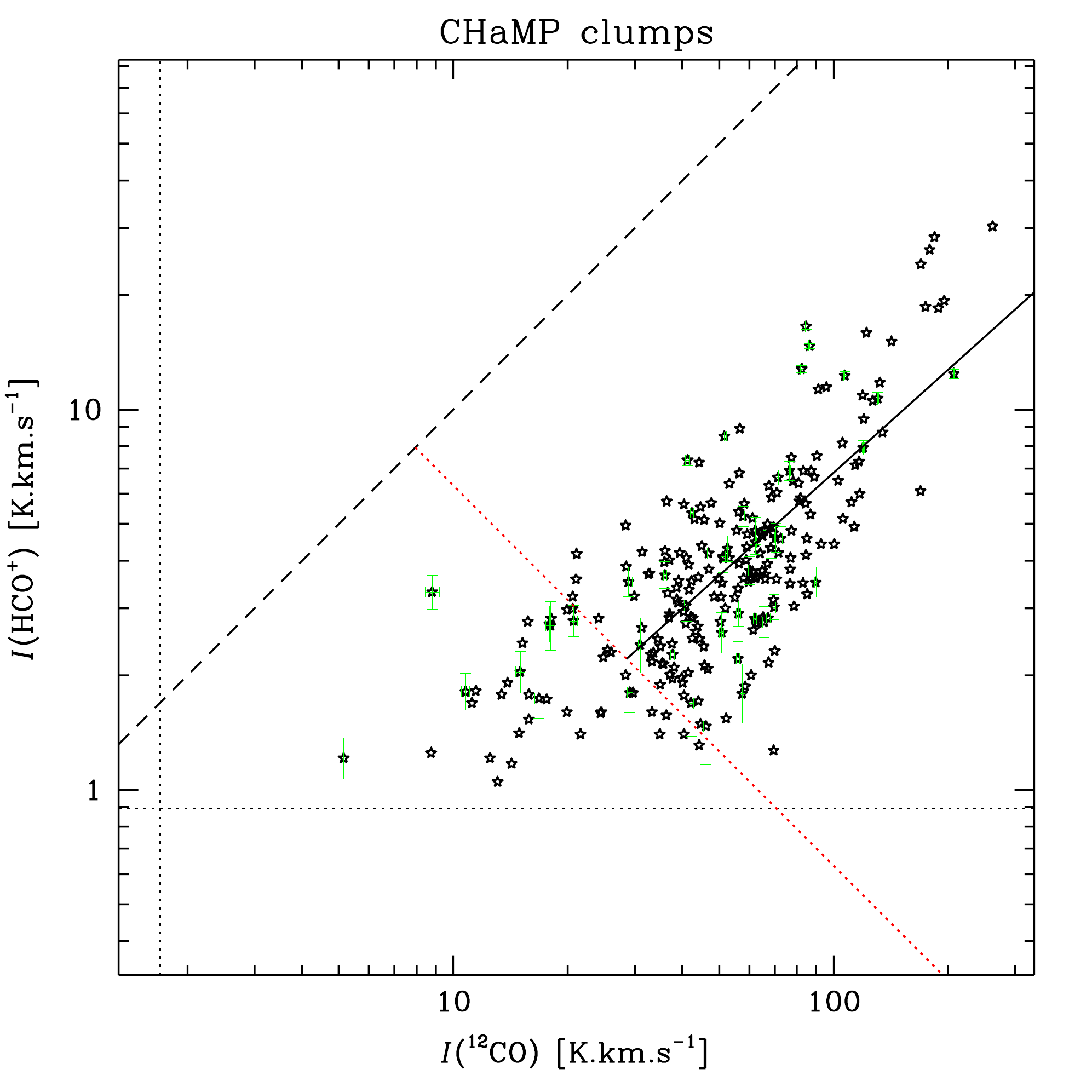}
\caption{Histogram of differences in the orientation (position angles) of CHaMP clumps' ellipses, as defined by \tco\ and \hcop\ emission.  Overlaid is a fit comprising a gaussian with a zero-level offset, with mean $\pm$ dispersion in the $\Delta$PA of -0\fdeg4 $\pm$ 14\fdeg5, and a reduced $\chi^2$ = 1.04.  (b) Comparison of integrated intensities between \hcop\ and \tco, with green error bars for uncertainties on 1-in-5 points, the mean 3$\sigma$ sensitivity limits as dotted lines, and a diagonal dashed line where both species would give the same intensities.  The solid line is a least-squares fit to those points above the dotted red line (which denotes a threshold where the \hcop\ data may be incomplete due to the sensitivity limit).  The solid line fit has slope 0.90$\pm$0.06, intercept -0.96$\pm$0.11, and correlation coefficient $r^2$ = 0.70.
}
\label{deltaPA}
\end{figure*}

\subsection{Comparison of Observed Properties Between Clump Envelopes and Interiors}\label{obscomps}

Next, it is instructive to directly compare the observed clump parameters (Table B1) from both the \tco\ and \hcop\ maps.  This is the kind of direct benefit we were aiming for when designing CHaMP: a multi-transition mapping study of a large sample of clouds, obtained with essentially the same angular and velocity resolution, sensitivity, uniformity, and completeness.  Such comparisons enable a systematic gauge of the significant trends among the clumps, without the extraneous observational factors that otherwise may exist when comparing disparate surveys, and which might skew any physical conclusions inferred from the data.

The \tco\ clumps' basic observable parameters include the peak brightness, linewidth (FWHM or dispersion), size at half-power, shape, and orientation.  As is evident from inspection of the maps, the morphology of the \tco\ clumps is (perhaps surprisingly, given the much higher optical depths) very similar in most locations to that of the \hcop\ maps (Fig.\,\ref{overlay}).  This similarity extends to the distributions of most of these basic parameters, and is quantified in Figures \ref{RDVcomps} and \ref{deltaPA}, where we see that the clump sizes, orientations, linewidths, and brightnesses are correlated in both species, despite their very different optical depths, abundances, and excitation conditions.

The clump sizes in particular (Fig.\,\ref{RDVcomps}a, where ``size'' is the geometric mean of the major and minor axes measured at half-power listed in Table B1) are well-correlated, with perhaps 70\% of the clumps being within 1$\sigma$ of the {\bf{\em same size}} in both species.  For the other $\sim$30\% of clumps (including the 3\% of blended structures, \S\ref{obspar}, which have the largest \tco\ sizes), they all show larger mean radii in \tco\ than in \hcop, with a maximum \tco/\hcop\ size ratio around 5. %
Thus, while both size distributions can be approximated by a gaussian function in log(size), and while both distributions have a minimum measurable size ($\sim30''$) set by the Mopra beam, the distribution for \tco\ is much broader, extending to sizes a few times larger than for \hcop.  Numerically, the mean $\pm$ SD of the \tco\ clump sizes (measured in log arcsec) is 2.10$\pm$0.25, as opposed to 1.96$\pm$0.18 for the \hcop\ clumps.  Measured in pc, the mean $\pm$ SD are $<$log\,$R_{\rm ^{12}CO}$$>$ = --0.08$\pm$0.27 and $<$log\,$R_{\rm HCO^+}$$>$ = --0.17$\pm$0.22.  Therefore, {\bf{\em on average}}, the \tco\ clumps are about twice as large in solid angle, $\sim$1.6$\times$ as large in physical area, and about twice as large in deprojected volume as the corresponding \hcop\ emission, when observed with Mopra's resolution and sensitivity.  So while the overall map morphology between these two species is quite similar, this result is a real difference in the emitting area visible in each clump from the two species.  Physically, this difference is to be expected, given the higher optical depth and lower excitation and column density requirements for seeing \tco\ emission, compared to \hcop\ emission.

Nevertheless, we maintain that this size difference is relatively ``small'', in the following sense.  Part of the received wisdom about molecular clouds is that \tco\ best traces the larger, GMC-scale (\gapp 10\,pc) structure of the lower-density envelopes, whereas ``dense gas tracers'' like \hcop\ are different because they better follow the active star formation \citep{L10}.  Instead, in Paper I we confirmed that \hcop\ is strongly subthermally excited nearly everywhere it is seen, typically tracing gas at densities perhaps 2 orders of magnitude below its critical density for thermalisation, as previously shown by \citet{e99} and others.  What is quite remarkable here is that \tco\ is largely tracing the {\bf{\em same structures}} as the \hcop, i.e., with sizes only a factor of 2 larger, rather than a factor of 10 or more.  In other words, at this resolution and sensitivity, the \tco\ emission morphology does not really define GMCs at all, even in large complexes, but rather is broken up into clump-sized ($\sim$1\,pc) units which contain most of the mass of even the large complexes and ``GMCs'' (we quantify this argument in \S\ref{clumpcomps}).

\notetoeditor{}
\begin{figure}[t]
\includegraphics[angle=0,scale=0.45]{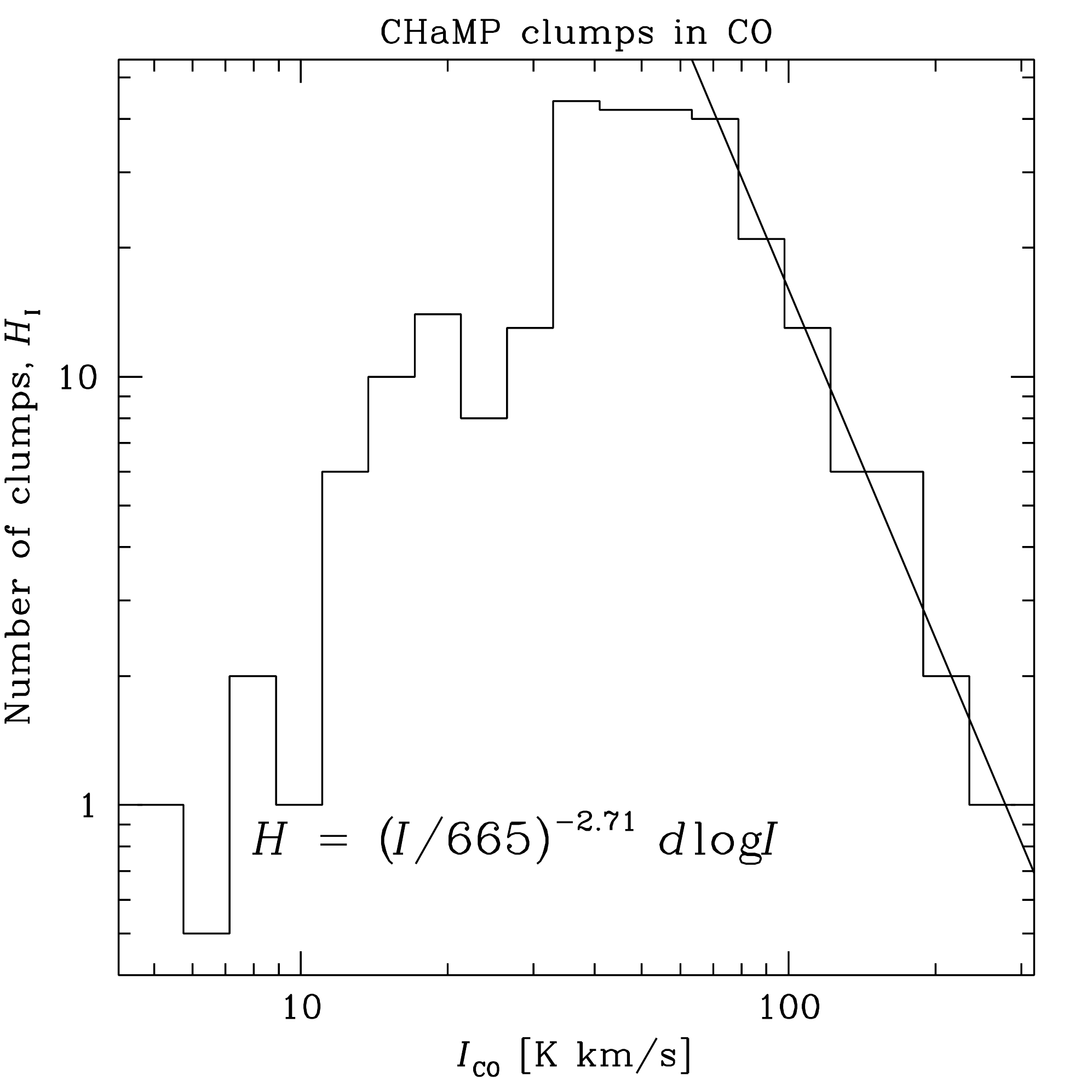}
\vspace{-4mm}

\caption{The \tco\ source function for the CHaMP clumps, shown here for 19 histogram bins, with a least-squares fit to bins \ico\ $>$ 60\,K\,\kms.  A maximum likelihood fit to the actual data in the same range gives values $I_0$ = 620$\pm$50\,K\kms\ and $p$ = --2.78$\pm$0.27, which is statistically preferred over any histogram fit.
This is a very different distribution from the \hcop\ source function in Paper I, since here the break in the power law below $\sim$60\,K\,\kms\ is around 20$\times$ brighter than the 5$\sigma$ sensitivity limit, whereas for the \hcop\ data, the break in the power law is only $\sim3\times$ the 5$\sigma$ sensitivity limit.  See text for further discussion.
}
\label{srcfn}
\end{figure}

Apart from the clumps' apparent sizes, the velocity dispersions in the two species can be thought of as a significant measure of the internal dynamical state of the clumps' envelopes and interiors.  We see in Figure \ref{RDVcomps}b that these are also broadly correlated around a mean ratio of 1, and a standard deviation of 0.5, with no clear asymmetry in the distribution towards larger \tco\ or \hcop\ linewidths.  However, there are a small number of outlier clumps ($\sim$5\%) which have linewidth ratios $>$3, but not in favour of either species.

The  similarities between these two species also extend to the clump shapes and orientations (Fig.\,\ref{deltaPA}a).  While a fraction of the clumps appear to have random orientations between the two species (as defined by the respective major axis PAs), most are strongly aligned, with a mean $\pm$ SD difference of only $\Delta$PA = --0\fdeg4 $\pm$ 14\fdeg5.  In other words, $>$70\% of the clumps have their emission aligned to $<$30\degree\ for the two species.  Likewise, the \tco\ clump shapes (defined as the aspect ratios of the major to minor axes) have mean $\pm$ SD = 2.00 $\pm$ 0.94 and median $\pm$ SIQR = 1.81 $\pm$ 0.46, statistically indistinguishable from the \hcop\ values.  The orientations with respect to the Galactic plane are equally random in both species.  And the distribution of peak \tco\ brightness seems unrelated to either the \tco\ size or linewidth, as was found for \hcop.  Because of these similarities, we do not show the corresponding \tco\ plots here.

But the two species' brightnesses are also strongly correlated with each other (Fig.\,\ref{deltaPA}b), with mean $\pm$ SD ratios \tbtco/\thcop\ = 7$\pm$2 and \ico/\ihcop = 10$\pm$2.  It seems inescapable that the clumps' envelope properties are strongly tied to those of their denser interiors.

\notetoeditor{}
\begin{figure}[t]
\includegraphics[angle=0,scale=0.45]{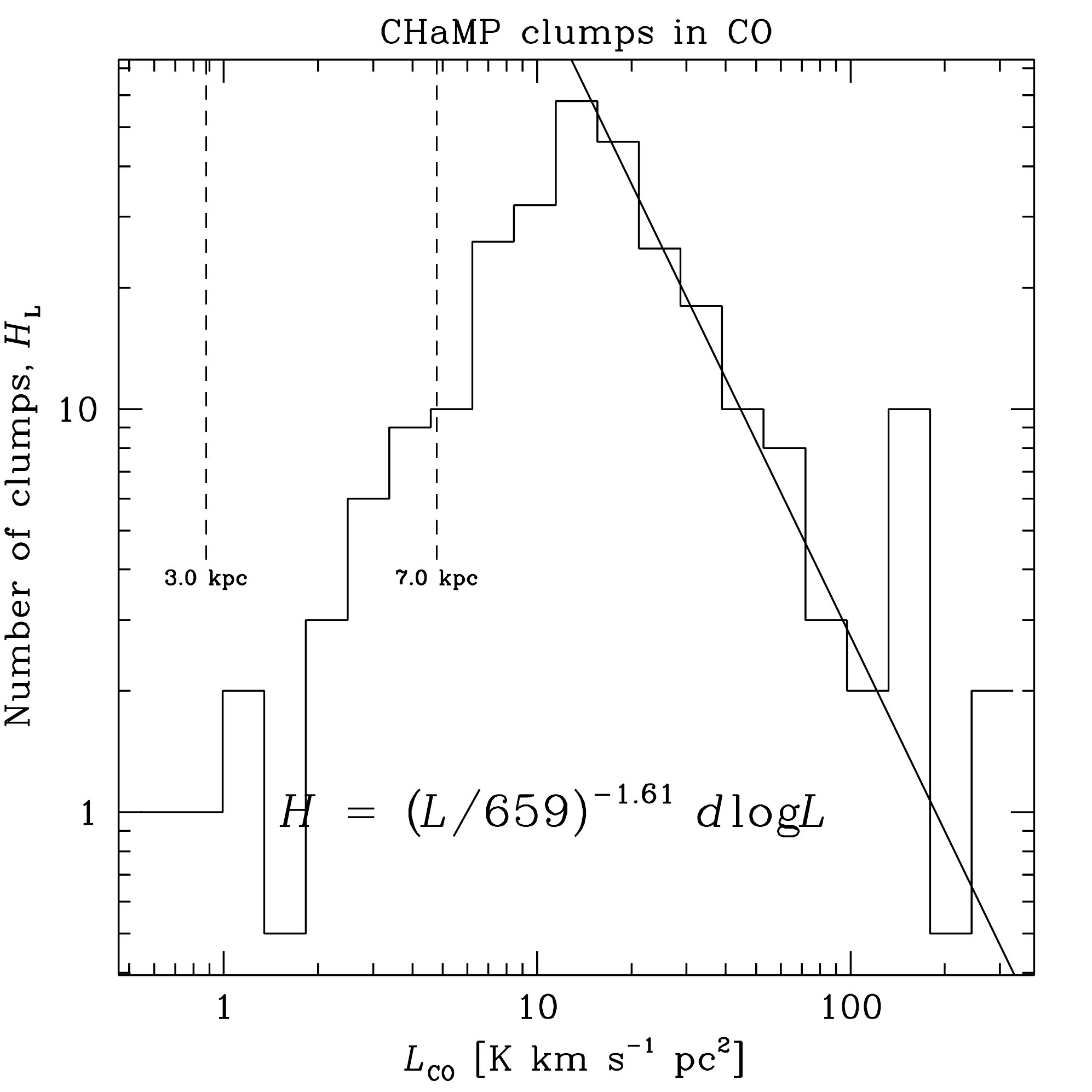}
\vspace{-4mm}

\caption{The \tco\ line luminosity PDF for the CHaMP clumps, shown for 21 histogram bins, with a least-squares fit to bins 11.5\,K\,\kms\,pc$^2$ $<$ \ico\ $<$ 123\,K\,\kms\,pc$^2$.  A maximum likelihood fit to the actual data in the same range gives values $L_0$ = 670$\pm$100\,K\kms\,pc$^2$ and $q$ = --1.60$\pm$0.12.  The dotted vertical lines show our sensitivity limits at the two indicated distances.  See text for further discussion.
}
\label{lumfn}
\end{figure}

\subsection{Distributions of Clump Envelope Properties}\label{coplots}

We now examine the distribution of various cloud parameters as measured by the \tco\ emission.  The first statistic we examine is the source probability density function (PDF), or source function.

Paper I made the case that the sample of \hcop\ clumps was essentially complete down to an integrated intensity of $\sim$4\,K\kms, based on the appearance of that source function.  In principle, we could search the \tco\ maps presented here and expect to reach an equivalent (allowing for the higher noise level) completeness limit near $\sim$8\,K\kms.  Above this limit, a complete sample would be expected to show a power-law distribution up to the brightest cloud in the sample.  However, the original Nanten Master Catalogue was based on an unbiased selection of \hcop- and \ceto-emitting clouds, {\bf{\em not}} on the \tco\ emission.  Therefore, our mapped areas will not necessarily be an unbiased sample of all molecular gas traced by \tco.  This is why we have deliberately selected for analysis here, even within the maps we have, only those \tco\ clumps that are detectable in the \hcop\ maps and whose properties have been compiled in Paper I.  Thus, we expect this sample of clouds will be far from complete in terms of their \tco\ properties alone.  But we assert that the properties described herein {\bf{\em will}} be fully representative of the envelopes of the denser gas clumps delineated in \hcop.

\notetoeditor{}
\begin{figure}[t]
\includegraphics[angle=0,scale=0.45]{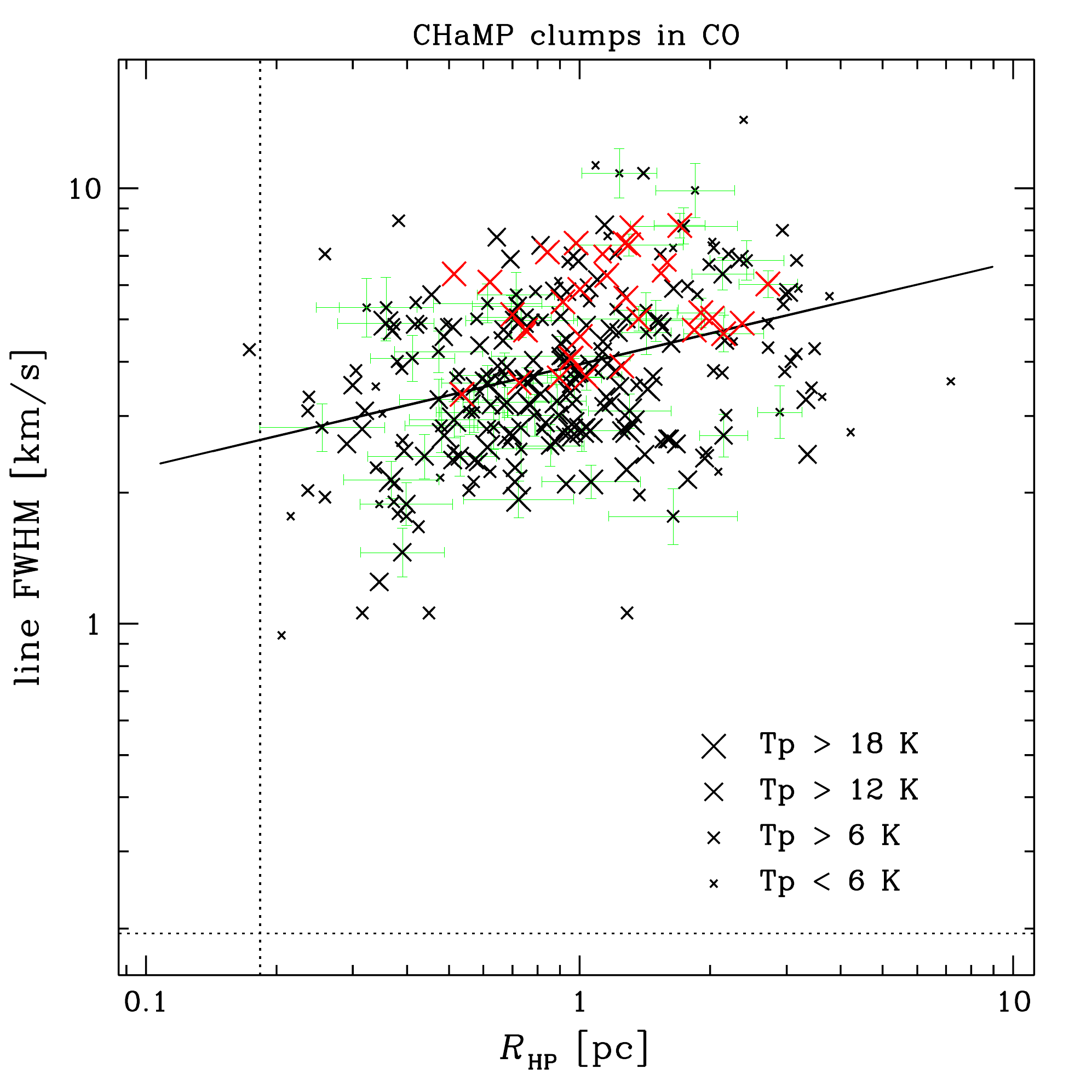}
\vspace{-4mm}

\caption{Size-linewidth relation for Mopra \tco\ clumps. %
The symbol size corresponds roughly to the magnitude of the peak temperature. %
Uncertainties for one in five points are shown as green error bars and mean sensitivities as dotted lines. %
The red symbols show the brightest clumps (\ico\ $>$ 90\,K\kms). %
The black line shows a least-squares fit to all points, corresponding to %
$\Delta$V(FWHM) = (3.94$\pm$0.10\,\kms)\,$R^{0.24\pm0.04}$ with an rms scatter of 0.18 in the log (a factor of 1.5), and a correlation coefficient $r$ = 0.59.
}
\vspace{-1mm}
\label{sizelw}
\end{figure}

This distinction is evident in Figure \ref{srcfn}, where there is no single power-law
\begin{equation}   
	H_W = (I/I_0)^p~d\,{\rm log}\,I~
\end{equation}
above 8\,K\kms, only above $\sim$60\,K\kms.  The break to much smaller numbers of clumps in the range 8\,K\kms\ $<$ \ico\ $<$ 60\,K\kms\ is clearly a selection effect of being concerned here with only those \tco\ clumps capable of producing detectable \hcop\ emission, and so requiring higher column densities and/or excitation conditions than the bulk of fainter \tco\ clouds that do not have associated detectable \hcop.  %

We also examine the \tco\ line luminosity PDF, %
\begin{equation}   
	H_L = (L/L_0)^q~d\,{\rm log}\,L~,
\end{equation}
one example of which is shown in Figure \ref{lumfn}.  %
This fit is over the luminosity range 11.5\,K\kms\,pc$^2$ $<$ \lco\ $<$ 123\,K\kms\,pc$^2$, below which we have the expected deficit of fainter clouds (but well above the sensitivity limit) due to our selection procedure.  However, above this range there appears to be a slight excess of bright sources above the fitted power-law, 3 expected vs.\,12 observed.  A 2-sided KS test reveals that this difference is statistically marginal (33\% chance that the deviation is drawn from a different population), but if this difference were real, it might be related to the distinct subsample (5\%) of ``\hcop-bright clumps'' from Paper I.

Similarly, we can calculate the \tco\ mass PDF, but because of the selection effects described above, this does not contain any useful information, and we do not discuss it further in this paper.

\notetoeditor{}
\begin{figure}[t]
\includegraphics[angle=0,scale=0.43]{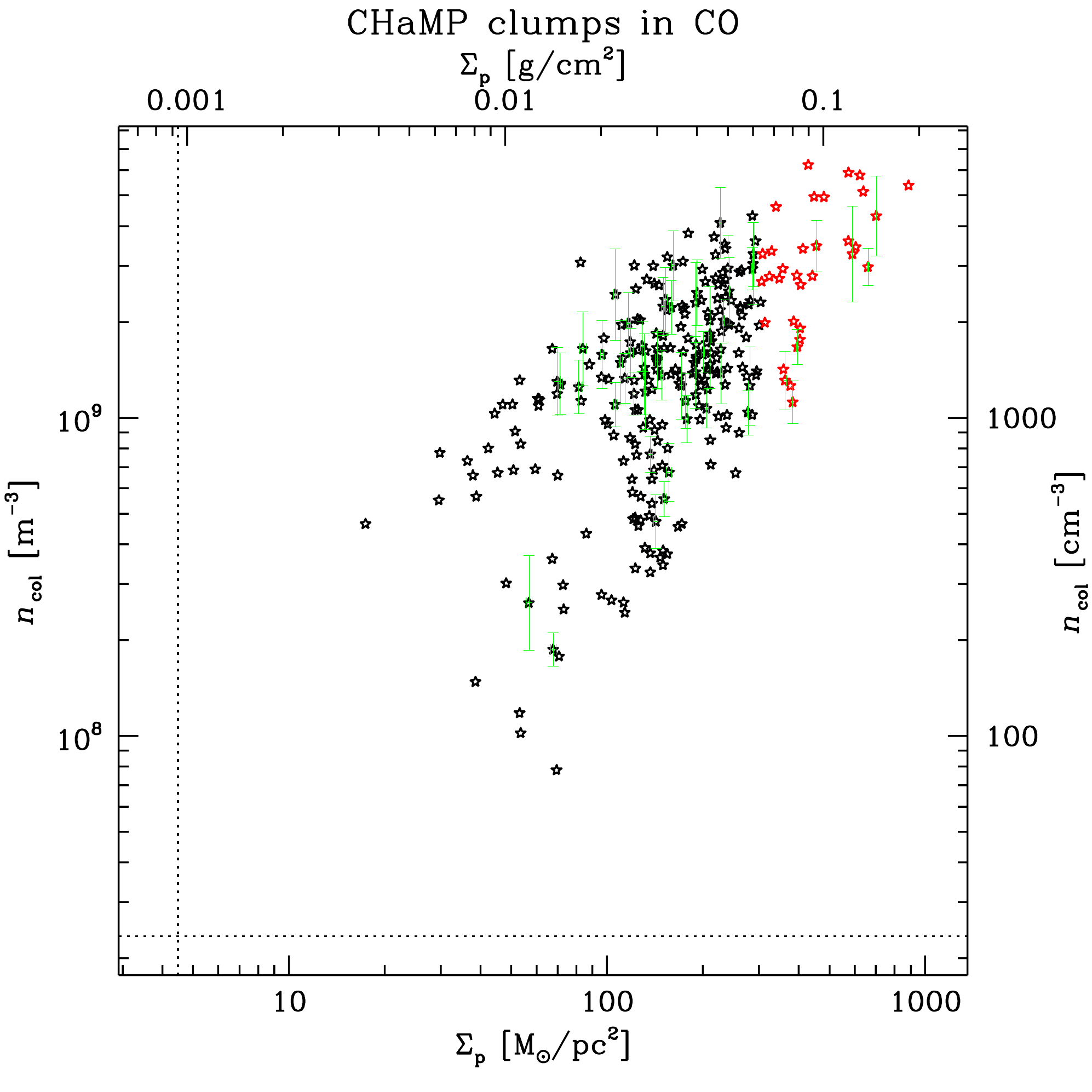}
\vspace{-4mm}

\caption{Volume density vs.\ \tco\ \joz\ mass surface density (proportional to column density) for CHaMP clumps, using the \tco\ $T_p$ values to determine the \tex. For each axis, we provide a natural and cgs scale for convenience. Uncertainties, 3$\sigma$ sensitivities, and the high-brightness tail of the source PDF are also shown as in Figure \ref{sizelw}. Since $n\propto(N/R)$, the scatter in this plot is due entirely to the clump radius.
}
\label{zigdens}
\end{figure}

Perhaps more interestingly, in \tco\ we find a distinct size-linewidth relation, albeit with a large scatter, as shown in Figure \ref{sizelw}.  This is unlike the much weaker size-linewidth relation found among the \hcop\ clumps.  The index we find here for the \tco\ size-linewidth relation, 0.24$\pm$0.04, is similar to, but somewhat smaller than, a number of other studies of the ``Larson relations'' in molecular clouds \citep[][and references therein]{hd15}, and is generally attributed to these clouds being turbulent structures.  The fact that the size-linewidth relation is much weaker in the denser gas traced by \hcop\ (Paper I, where we obtained a fitted index 0.12$\pm$0.05) suggests in contrast that turbulence does not act alone in determining the properties of the dense gas.  Especially for the lower-mass clumps, most cannot be in virial equilibrium unless confined by an external pressure.  We discuss this topic further below.

\begin{figure}[t]
\includegraphics[angle=0,scale=0.47]{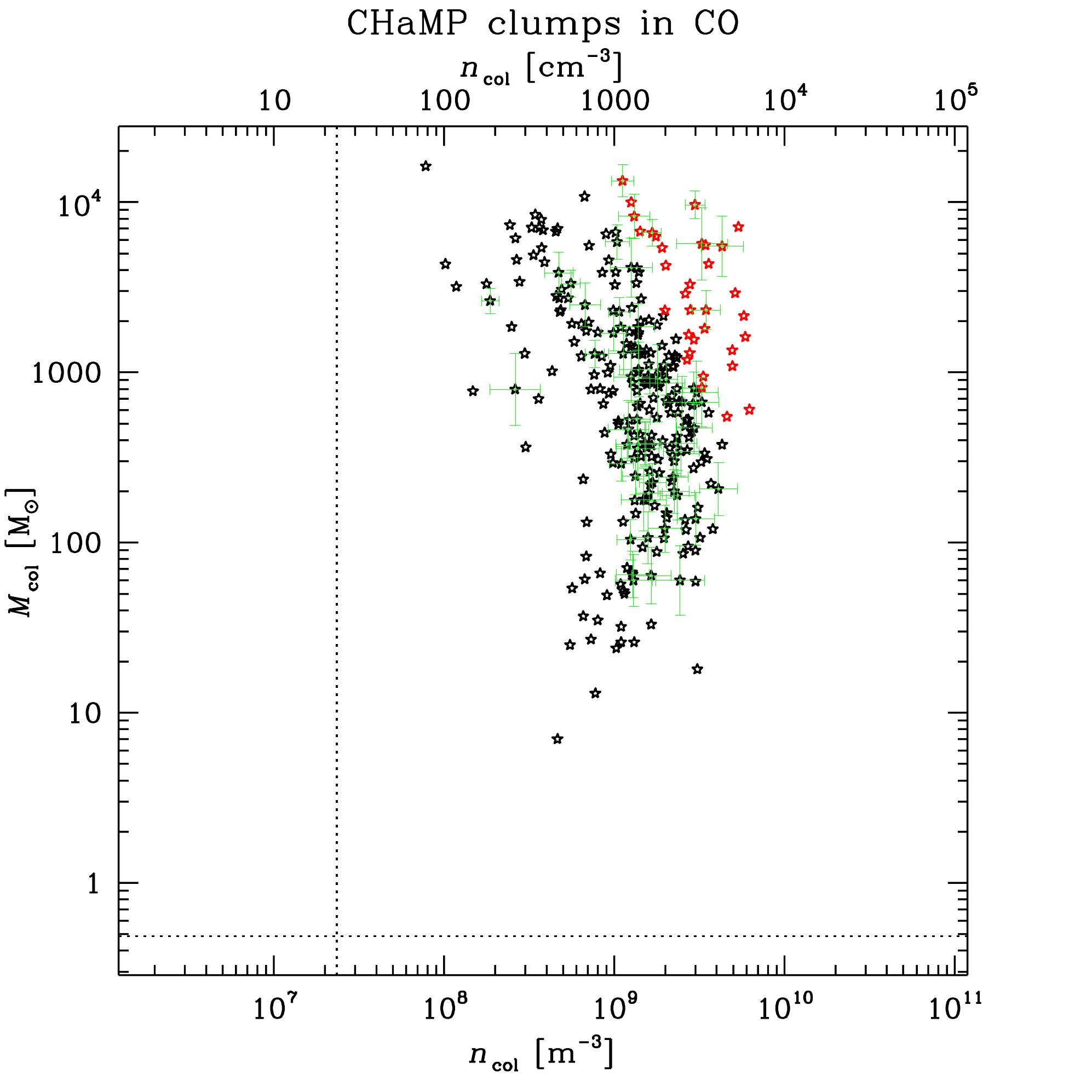}
\vspace{-5mm}

\caption{Mass from column density vs. volume density for Mopra \tco\ clumps. Other details, including 3$\sigma$ sensitivity limits, are as in Figure \ref{sizelw}.
}
\vspace{-1mm}
\label{Mdens}
\end{figure}

Because we use the $X$ factor to convert \ico\ to column density (eq.\,4) or mass surface density (eq.\,5), the distribution of these quantities will be the same as shown in Figure \ref{srcfn} for the integrated intensity.  Neither do we have a distribution of optical depths to examine here.  We can, however, use the measured clump sizes to convert the column densities to volume densities (Fig.\,\ref{zigdens}) and to clump masses (Fig.\,\ref{Mdens}), as was done in Paper I (see there for the procedures and formulae used, for Figs.\,\ref{zigdens}{\em ff}).

These plots reveal that, apart from the density, the clump envelopes (which we take to be well-traced by the \tco\ emission) possess similar, but not identical, bulk properties to their denser interiors (for which we take the \hcop\ parameters to be typical).  Thus, the peak column or surface densities  of the envelopes/interiors range in log(M\solar\,pc$^{-2}$) over roughly 1.5--3.0/1.5--3.5, respectively.  The masses range approximately over 1.0--4.0 in log(M\solar) for both.  These ranges are (perhaps surprisingly) quite similar for the two species.  But as might be expected from the molecules' different excitation requirements and abundances, the inferred peak densities are different, ranging roughly over 7.9--9.8/8.3--10.5 in log(m$^{-3}$) for the envelopes and interiors, respectively.  These results suggest that the two species trace similarly-massed clumps in molecular clouds, but that \hcop\ preferentially maps the interiors up to $\sim$2$\times$ higher column density, or over $\sim$3--5$\times$ higher volume density, than \tco, while \tco\ more readily maps $\sim$2$\times$ the area of the \hcop\ clumps towards their outer peripheries.

\notetoeditor{}

We note an additional difference between the \tco\ mass-density distribution (Fig.\,\ref{Mdens}) and the equivalent plot for \hcop.  In Paper I (Fig.\,15), we found that such a plot illustrated the difference most clearly between the ``bright clump'' subsample and the rest of the \hcop\ clumps.  There, the 5\% of bright clumps were found to have systematically higher densities and masses than the rest of the sample.  We further showed \citep{b13} that this difference was attributable to these clumps' higher star formation activity, as measured by the \brg\ emission associated with each clump.  Here, however, we cannot discern an equivalent distinction among the brightest \tco\ emitters; this may be partially due to the way we have converted \ico\ to column density using the $X$ factor, possibly underestimating the column or volume density in the highest optical depth clumps.  We suppose that once the \tco\ optical depths are obtained, we might recover this distinction. %

This last point is important: the standard $X$-factor approach used so far for converting \ico\ to \nhtwo\ does not explicitly take into account the \tco\ optical depth, as mentioned in \S\ref{physpar}.  We may well find different results if we could obtain and use such information, but as we shall see in \S\ref{newx}, new insights from other studies may enable a parametrised way to improve this conversion.

The significance of the \ico\ to \nhtwo\ conversion can be understood with the following realisation: while the overall \hcop\ and \tco\ density and mass ranges may seem superficially reasonable and consistent, they actually hint at a possible discrepancy.  If the \tco\ is at least approximately tracing the molecular cloud column via the $X$ factor, then the mass of the envelopes of the \hcop\ clumps they are expected to be tracing (as indicated by the larger mapped clump sizes) should {\bf{\em contain}} the mass of the denser interiors.  In other words, if the $X$ factor prescription were correct as widely used in the literature, the \tco\ maps should integrate to {\bf{\em at least}} the same \hcop\ masses over the same areas as the \hcop\ clumps, and to a {\bf{\em larger}} mass when integrated over the larger areas of the envelopes.

\notetoeditor{}
\begin{figure}[t]
\includegraphics[angle=0,scale=0.45]{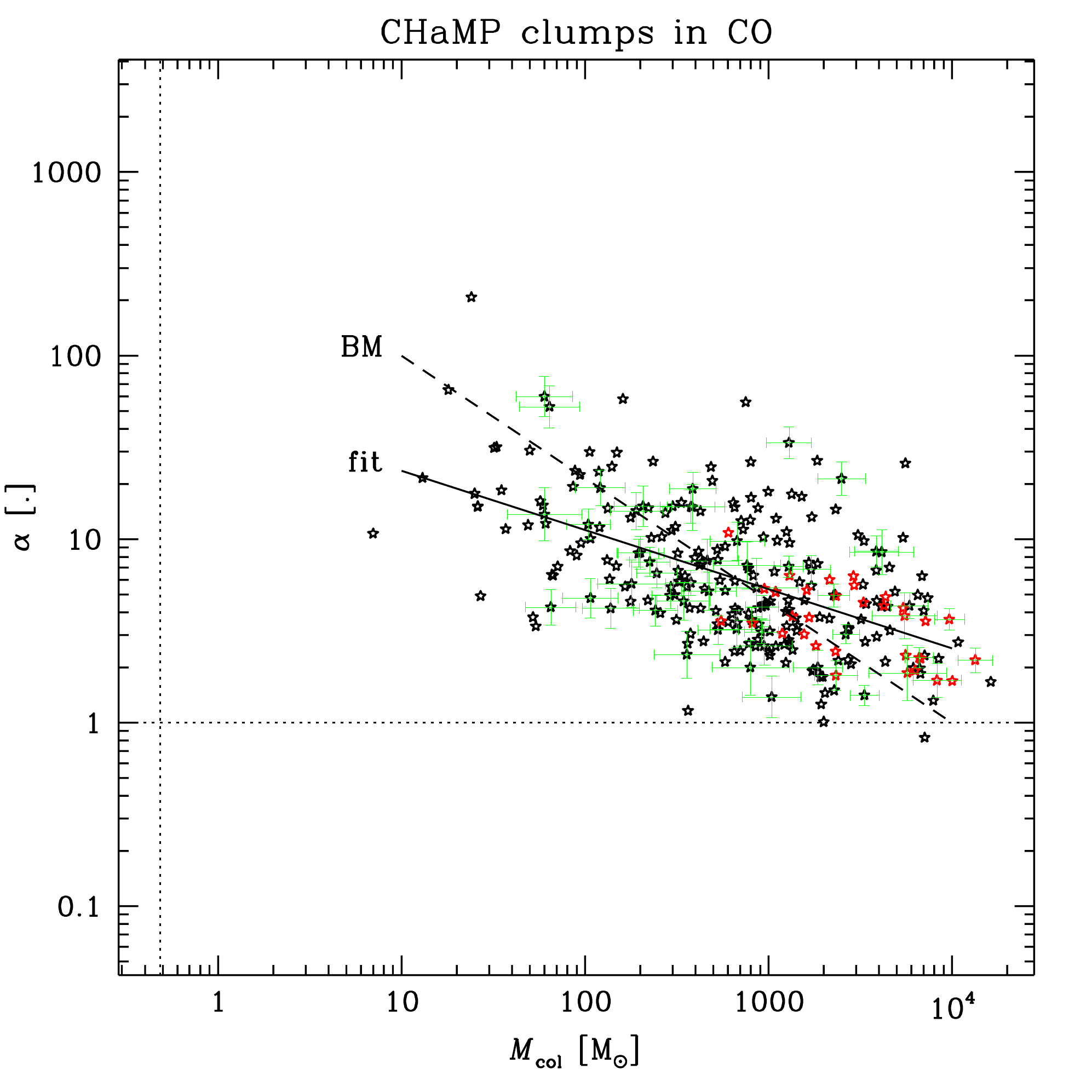}
\vspace{-4mm}

\caption{\citet{bm92} $\alpha$ parameter vs. mass from column density for Mopra \tco\ clumps. The solid line in each panel (labeled ÒfitÓ) is the least-squares fit to the clump data, while the dashed line (labeled ÒBMÓ) is the theoretical line from Bertoldi \& McKee (1992). The dotted lines show the 3$\sigma$ mass sensitivity and the $\alpha$ = 1 limit for gravitationally supported clouds. Other details are as in Figure \ref{sizelw}.
}
\label{alpha}
\end{figure}

The fact that the clumps seem to mass similarly in both tracers over different areas suggests that either (a) the \hcop\ masses in Paper I are systematically too high (e.g., due to the assumed value for $X_{\rm HCO^+}$ being too small), or (b) the \xco\ factor used here should be larger at higher columns ($\Sigma$ \gapp\ 300\,M\solar\,pc$^{-2}$) or densities ($n\gapp10^{9.5}$\,m$^{-3}$).  Based on the discussion of abundances in Paper I as well as more recent results \citep[e.g.,][]{gbs14}, we discount possibility (a), and in this paper focus instead on option (b).  A variable \xco\ could be due to a number of factors which we discuss in the next section.

Another manifestation of this issue is the distribution of virial masses compared to the masses derived from the column densities, or virial-$\alpha$ \citep{bm92,kpg13}.  We show this in Figure \ref{alpha}, where the general pattern is very similar to the distribution for the denser interiors as measured in Paper I.  That is, $\alpha$ trends from $\sim$2 at the higher clump masses to several tens for the lower clump masses.  Again, this is potentially at odds with expectations that the $X$-factor requires $\alpha$$\sim$1: we discuss this further in \S\ref{clumpcomps}.

\notetoeditor{}
\begin{figure}[t]
\vspace{-5mm}

\includegraphics[angle=0,scale=0.43]{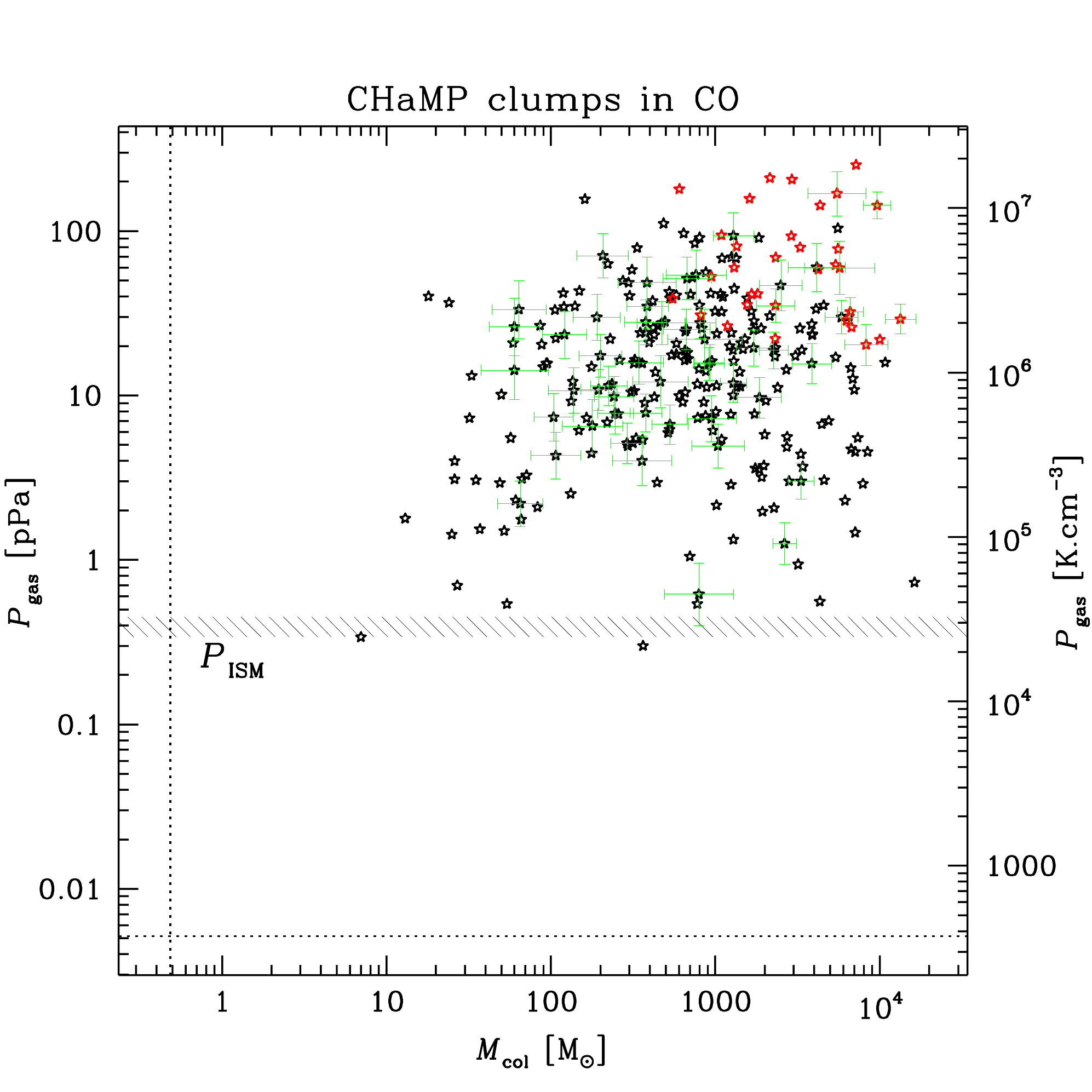}
\vspace{-4mm}

\caption{Total internal gas pressure vs.\ mass from column density for Mopra \tco\ clumps. The horizontal shaded region shows the level of the general ISM pressure from \citet{bc90}. Other details are as in Figure \ref{sizelw}.
}
\label{pres}
\vspace{-2mm}
\end{figure}

As with the \hcop\ results, the least-squares fit to the trend (power-law index --0.32$\pm$0.03, almost identical to the \hcop\-derived index) is also less than the theoretically expected value of $-\frac{2}{3}$.  %
Likewise, a plot of the ratio of \tco\ clump mass to Bonnor-Ebert mass looks very similar to that presented in Paper I.

In contrast, a plot of total internal pressure (thermal $+$ nonthermal) vs.\ mass, as computed for the envelopes and shown in Figure \ref{pres}, is noticeably different to the equivalent plot for the dense interiors in Paper I. Here we use \tex\ from the \tco\ data for each clump to compute the thermal contribution to $P_{\rm gas}$, compared to a constant \tex\ = 10\,K for the \hcop.  However, in both cases the nonthermal contribution dominates $P_{\rm gas}$.  While both distributions reach a floor near the value for the general pressure of the ISM, the \tco\ envelopes are significantly underpressured at their maximum values, by a factor of $\sim$3, compared to the maximum \hcop-derived pressures.  This result follows naturally from simply considering the envelopes' larger sizes and lower densities, but again, is subtly at odds with the expectation that the \tco\ data should trace a larger gas column than the \hcop.  We discuss this issue in more detail in \S\ref{clumpcomps}.

\notetoeditor{}
\begin{figure}[t]
\includegraphics[angle=0,scale=0.47]{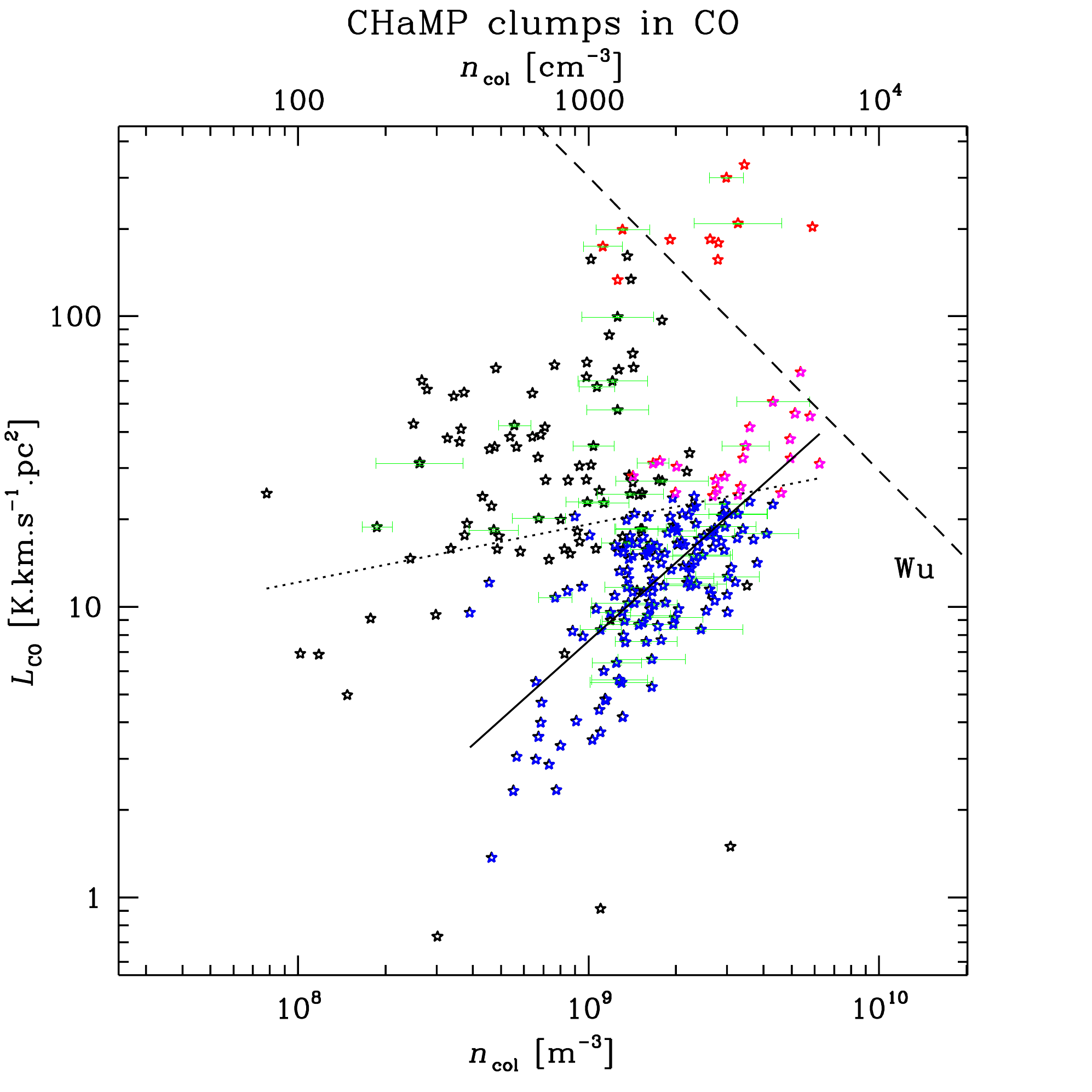}
\vspace{-4mm}

\caption{Mopra \tco\ integrated line luminosity vs.\ volume density, with weaker and brighter clumps from Figure \ref{srcfn} in black and red as before, except for 191 clumps with distance 2.4\,kpc $\leq d \leq$ 2.5\,kpc, which are shown in blue and magenta, respectively. We show two fits to these points: (1) all points were binned in equal intervals of log $n$ and these bins equally weighted in a least-squares fit, giving a power-law index 0.20 $\pm$ 0.09 (dotted line) and (2) a similar fit to the blue and magenta points, giving a power-law index 0.90 $\pm$ 0.06 (solid line). We also show, as a dashed line labeled ``Wu,'' the trend of \citet{wu10} on the same scale, as described in Paper I. %
}
\label{Ln}
\end{figure}

We conclude this section with a plot of the \tco\ line luminosity vs.\ volume density (Fig.\,\ref{Ln}), which is related to the Kennicutt-Schmidt star formation law, as explained in Paper I.  Briefly, the overall star formation rate (SFR) in a population of molecular clouds in a disk galaxy is thought to be adequately traced by the bolometric IR luminosity $L_{\rm IR}$, and is measured to be proportional to a power $N$ = 1.4--1.6 of the gas density $n_{\rm gas}$. %
The molecular line luminosity $L_{\rm mol}$ is also widely used as a proxy to measure the SFR \citep[e.g., as in][]{wu10}.  According to radiative transfer models \citep{kt07,n08}, $L_{\rm IR}$ $\propto$ $L_{\rm mol}^{a}$ and $L_{\rm mol}$ $\propto$ $n_{\rm gas}^{b}$, whence $N$ = $ab$.  The specific values of $a$ and $b$ will then depend on the details of radiative transfer in different species within the cloud population; generally, for higher critical density species like HCN, $a$ $\sim$ 1 and $b$ $\sim$ 1.5, but for lower critical density species like CO or \hcop, $a$ $\sim$ 1.5 and $b$ $\sim$ 1.

\notetoeditor{}
\begin{figure*}[t]
(a)\hspace{-2mm}\includegraphics[angle=0,scale=0.43]{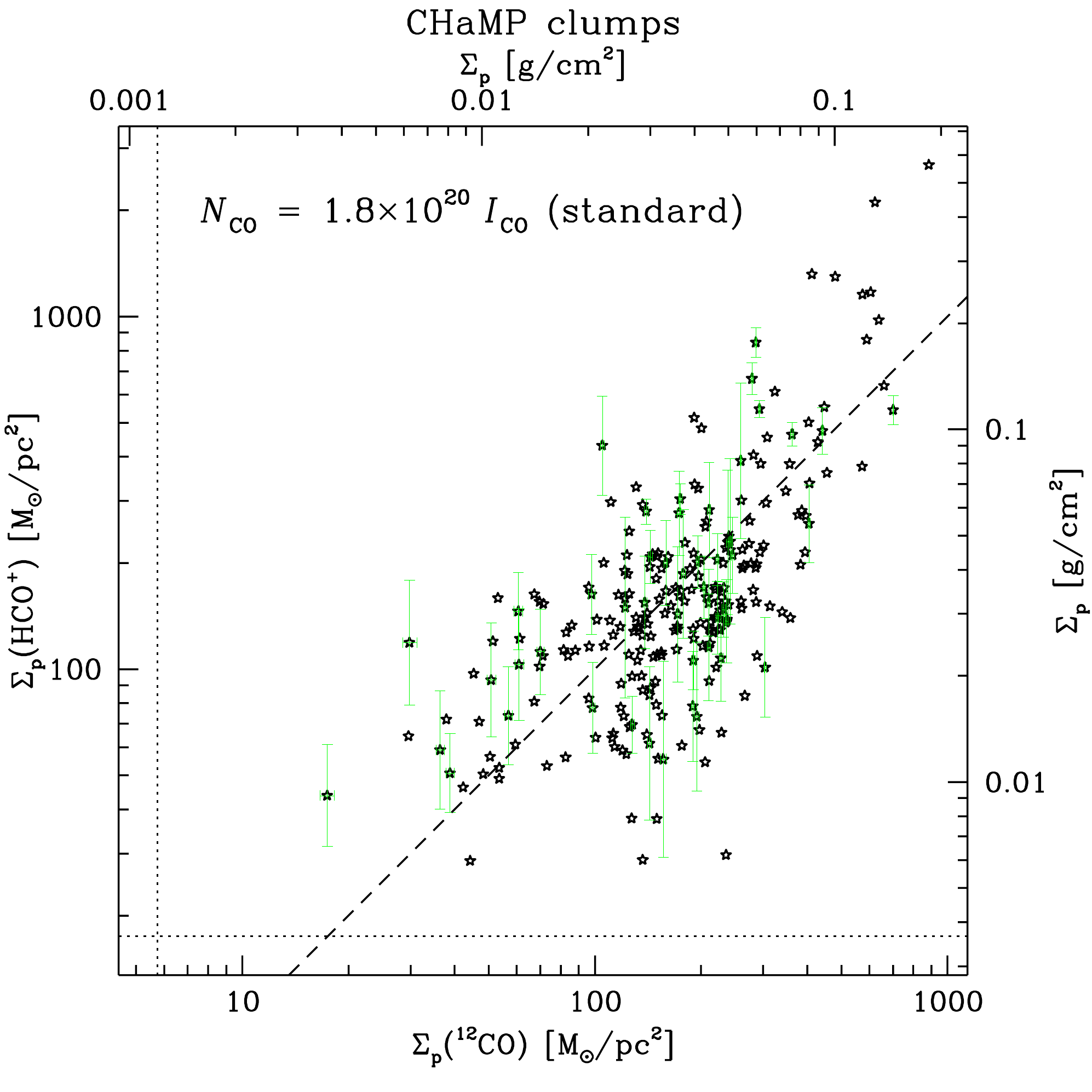}\hspace{2mm}
(b)\hspace{-2mm}\includegraphics[angle=0,scale=0.43]{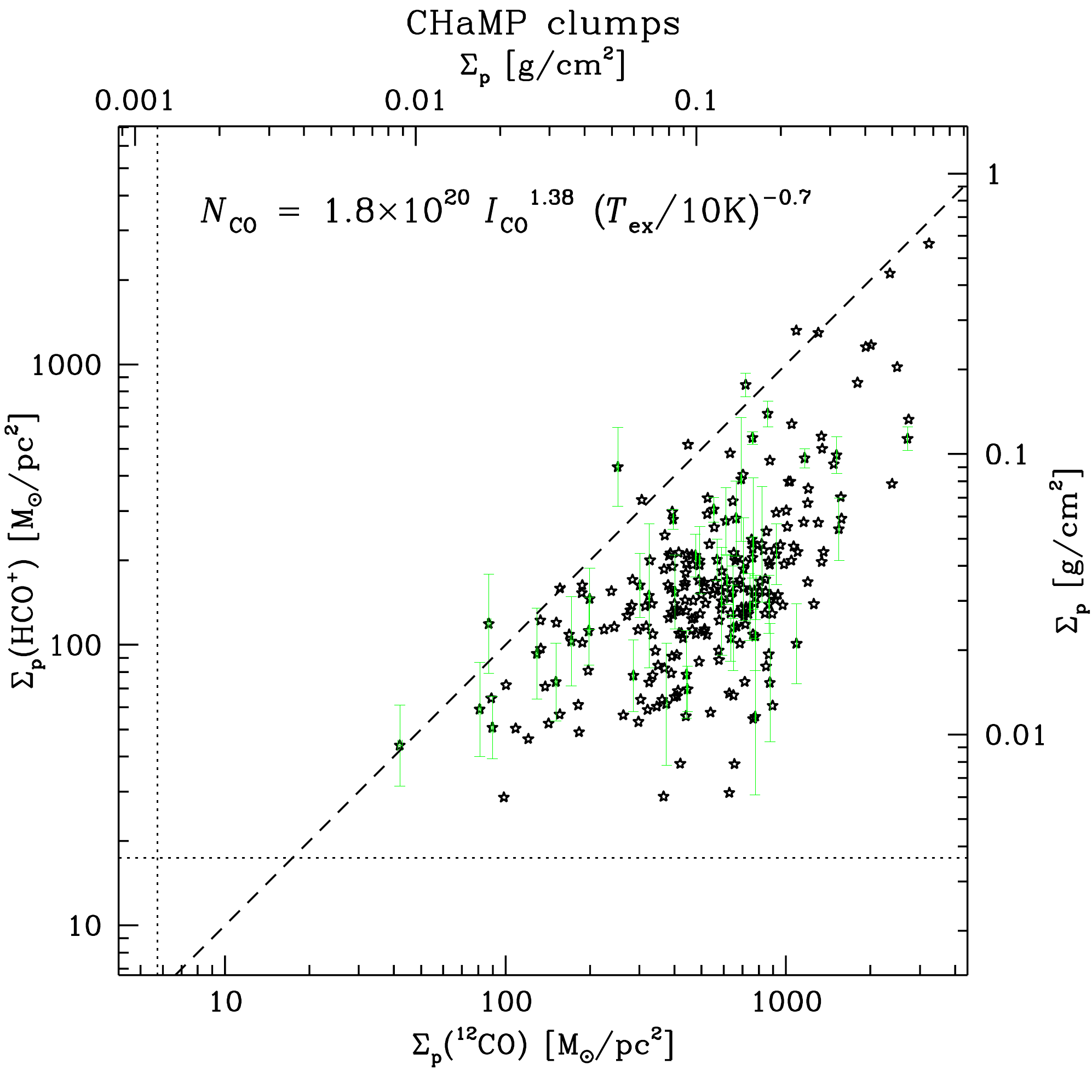}
\caption{Comparison of CHaMP clumps' mass surface density derived from \hcop\ measurements in Paper I ($y$-axis in both panels, assuming \xhcop\ = 10$^{-9}$ as in Paper I) and that derived from \tco\ measurements presented here ($x$-axis in both panels, assuming [\tco]/[\htwo] = 10$^{-4}$).  The left panel (a) shows $\Sigma$(\tco) using eq.\,(4), while the right panel (b) shows the same quantity but according to eq.\,(10), as labelled in each panel.  As in Fig.\,\ref{zigdens}, both panels show green error bars for uncertainties on 1-in-5 points, the 3$\sigma$ sensitivity limits as dotted lines, and a diagonal dashed line where both species would give the same $\Sigma$ with the assumed molecular abundances.
}
\label{Zcomps}
\end{figure*}

Therefore, Figure \ref{Ln} explores the value of the index $b$.  Confining ourselves to the blue and magenta points for the moment (those $\sim$60\% of clumps at a distance of 2.4--2.5\,kpc), the solid line shows a least-squares fit slope of $b$ = 0.90, while a robust fit gives an even steeper slope $b$ $\sim$ 1.3 (also suggested by the visual appearance).  This is much closer to \citet{n08}'s prediction of $b$ $\sim$ 1 than was the case for our \hcop\ data ($b$ = 0.44 observed vs.\ 1.0 predicted) on the same subsample.

This is partially due to the way we have calculated the volume density $n$, by using the \xco\ factor without calculating an optical depth.  In that case, $n$ $\propto$ \ico/$R$; while $L$ $\propto$ \ico\,$R^2$ $\propto$ $nR^3$.  This means that the clump subsample at distances near 2.5\,kpc is constrained to have $b$ = 1, plus an additional scatter from the clump size distribution (which is independent of \ico).  The other clumps in Figure \ref{Ln} at different distances (black/red points) then lie above the blue/magenta points because most of them are at larger distances, and for similar angular size and density distributions, their line luminosities $L$ $\propto$ $R^3$ $\propto$ $d^3$.  In contrast, for Paper I we calculated the \hcop\ volume densities via the optical depths, which introduces an additional scatter in the $n$ values, and makes the fitted slope shallower in those data.

\section{Analysis and Discussion}

\subsection{Alternative Column Density Conversions}\label{newx}

To compare cloud properties as derived from both \hcop\ and \tco, we first reconsider the standard approach \citep[eq.\,4 in \S\ref{physpar}, based on][]{dht01} to deriving physical parameters for molecular clouds from \tco\ data.  In particular, we focus on two recent studies where alternatives to this formula have been proposed.

As part of the ThrUMMS project, \citet{bm15} used radiative transfer analysis of iso-CO line ratio data over 120 deg$^2$ of the Fourth Quadrant of the Milky Way, to derive an intensity-dependent conversion,
\begin{equation}   
	N_{H_2} = 1.6\times10^{24}~{\rm H}_2\,{\rm mol\,m}^{-2}~(I_{\rm CO}/{\rm K\,km\,s}^{-1})^{1.38}~.
\end{equation}
Significantly, this approach converts the \tco\ data cubes directly into column density cubes with an implicit opacity correction, so the velocity-integrated parameters in Table B3 (such as mass) are moments of the $N$ cubes first, rather than using emission line velocity dispersions to calculate physical parameters, as for Table B2.  This formula results in column densities (and hence masses \& other quantities) that tend to be larger than those from eq.\,(4), especially for brighter regions with large \ico.

Similarly, as part of a project mapping iso-CO line emission and near-IR extinction across the California molecular cloud (CMC), \citet{kl15} showed that the conversion in this cloud is temperature-dependent,
\begin{equation}   
	N_{H_2} = 2.0\times10^{24}~{\rm H}_2\,{\rm mol\,m}^{-2}~\frac{(I_{\rm CO}/{\rm K\,km\,s}^{-1})}{(T_{\rm ex}/10\,K)^{0.7}}~.
\end{equation}
This formula will tend to produce column-related quantities which are smaller than those from eq.\,(4), especially for warmer regions with large \tex.

\notetoeditor{}
\begin{figure*}[t]
(a)\hspace{-2mm}\includegraphics[angle=0,scale=0.43]{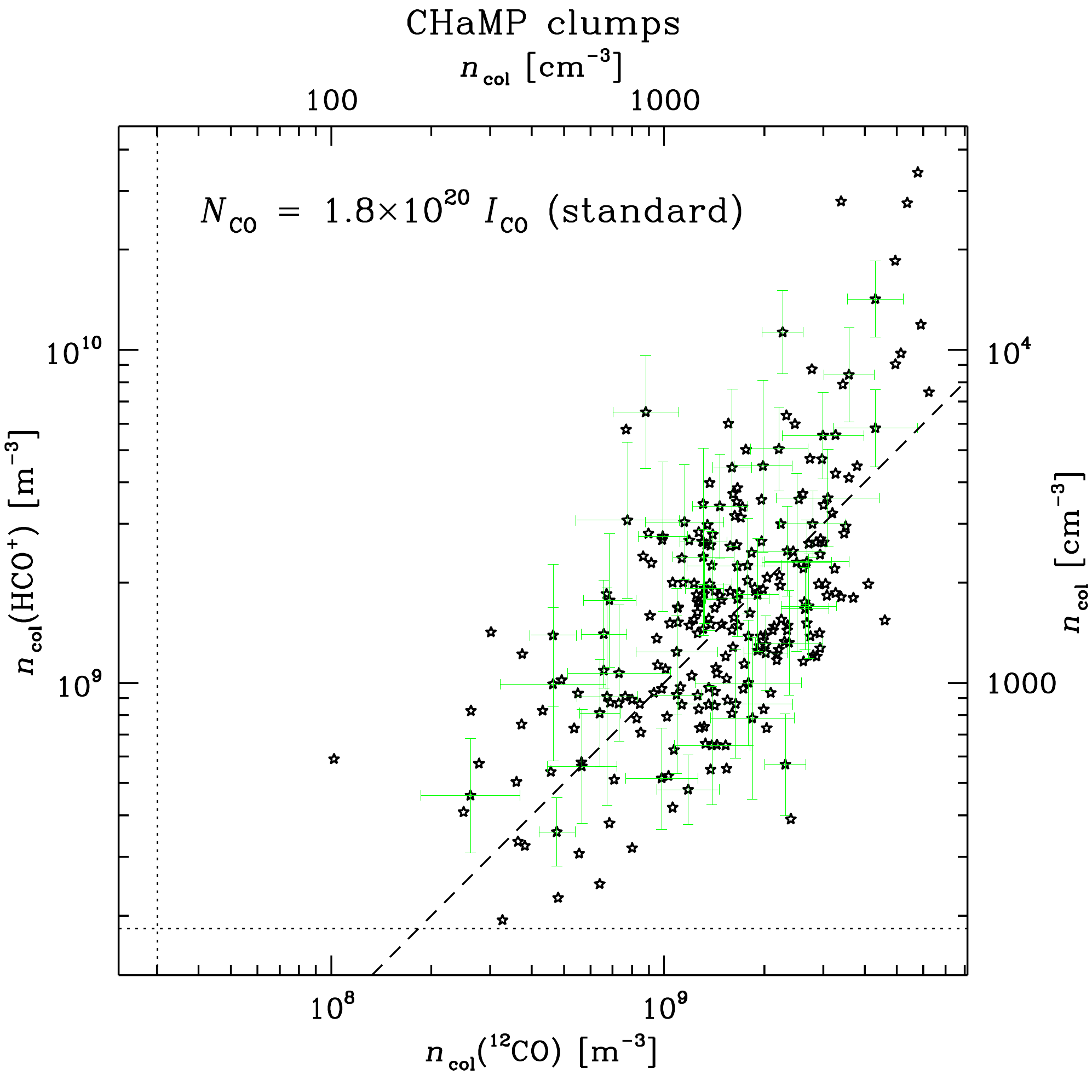}\hspace{2mm}
(b)\hspace{-2mm}\includegraphics[angle=0,scale=0.43]{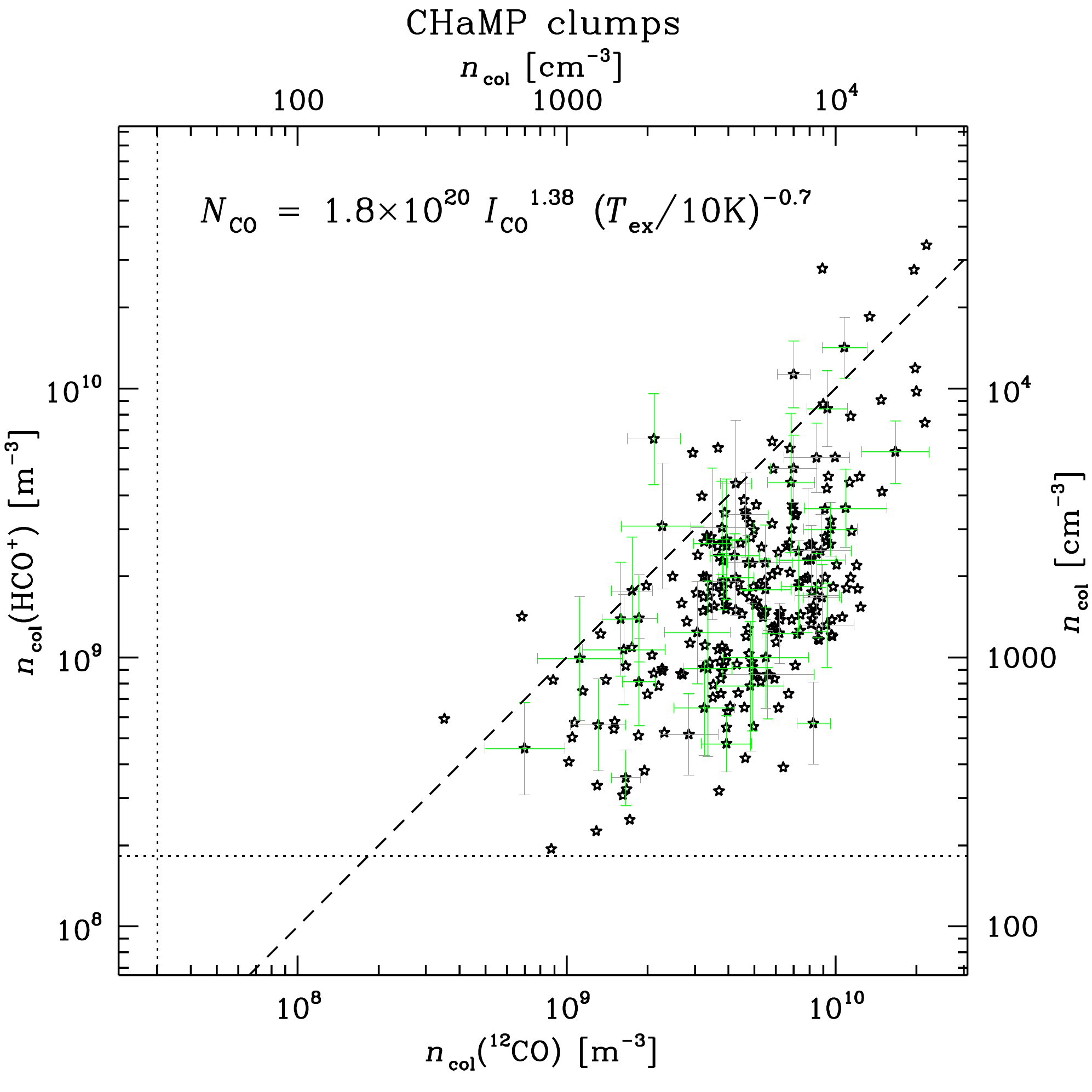}
\caption{Comparison of CHaMP clumps' gas number density derived from \hcop\ measurements in Paper I ($y$-axis in both panels, assuming \xhcop\ = 10$^{-9}$ as in Paper I) and that derived from \tco\ measurements presented here ($x$-axis in both panels, assuming [\tco]/[\htwo] = 10$^{-4}$).  The left panel (a) shows the implied $n$(\tco) using eq.\,(4), while the right panel (b) shows the same quantity but according to eq.\,(10), as labelled in each panel.  Other details are as in Fig.\,\ref{Zcomps}.
}
\label{ncomps}
\end{figure*}

Although the two new formulae might superficially seem to describe contradictory behaviour, they are actually complementary, since the exponent for \ico\ in eq.\,(8) derives mostly from a correction for high optical depth $\tau$ in the \tco\ line.  Further, besides yielding eq.\,(8), the ThrUMMS results simultaneously hint at a \tex-dependence that acts in the same sense as eq.\,(9).  Therefore, it is quite likely that both results are correct, providing a simple parametrisation for a more sophisticated conversion from \ico\ to \nhtwo.  Taking into account the respective normalisations, a combined formula is
\begin{equation}   
	N_{H_2} = 1.8\times10^{24}~{\rm H}_2\,{\rm mol\,m}^{-2}~\frac{(I_{\rm CO}/{\rm K\,km\,s}^{-1})^{1.38}}{(T_{\rm ex}/10\,K)^{0.7}}~.
\end{equation}
In practice, the numerical effect of eq.\,(8) is somewhat larger than that of eq.\,(9); however, both are significant.

Note that the 1$\sigma$ dispersion in the correlation that gives rise to eq.\,(8) is relatively small, $\pm$0.4--0.5 dex, and even less in the combined relation eq.\,(10), $\pm$0.25 dex for \ico\ $>$ 7\,K\,\kms\ (Barnes et al., in prep.), despite its appearance in Figure 15 of \citet{bm15}.  There, the contours of voxel incidence are in successive factors of 10, and are spaced much more widely than the statistical dispersion in any narrow bin of \ico.

It is important to understand the significance of this approach.  By using the isotopologue-calibrated conversions (eqs.\,8--10), which are derived over narrow 1\,\kms\ bins and do not depend on the actual linewidths, we implicitly allow for the high \tco\ optical depth and variable \tex.  Although a similar analysis to ThrUMMS for the iso-CO CHaMP data is also possible, at this point it is unnecessary, since the ThrUMMS sample of clouds is much larger than for CHaMP, and the overall calibration of these relations is unlikely to be different in the sample examined here.  This means that we can immediately perform a much more sophisticated analysis of \tco\ data alone than has been possible before.  Incidentally, this strategy is also applicable to other \tco\ projects by other workers.

We therefore re-derive all the \tco\ clump physical properties with eq.\,(10) instead of eq.\,(4), and provide these in Table B3 in the same format as Table B2.

\subsection{Comparison of Physical Parameters Between Clump Envelopes and Interiors}\label{clumpcomps}

One expects the clump properties as derived from the \tco\ emission, which is most likely tracing the lower-density envelopes of the clumps, to differ from the properties of the denser gas traced by the \hcop\ as compiled in Paper I.  Although such differences have been explored in a number of previous studies, we have the advantage of a large, uniform sample over which to make the comparisons.  As far as we are aware, the CHaMP clumps are the second-largest sample of molecular clouds mapped in \hcop\ alone \citep[after MALT90;][]{j13}, but the largest such sample with comparable maps in {\bf both} iso-CO and \hcop\ lines.  Thus, we should be able to differentiate the gas properties in the different emitting environments in a more statistically significant way than has been possible previously.

Therefore, we now compare in detail the \hcop\ results from Paper I with the current \tco\ data.  For completeness, we discuss comparisons using {\em both} eqs.\,(4) and (10).  The basic quantity in this discussion is the column density itself, or equivalently, the mass surface density $\Sigma$ (assuming an abundance relative to \htwo\ for each species, as in eq.\,5).  We show in Figure \ref{Zcomps} two comparisons, between the $\Sigma$ derived from the \hcop\ column density, and each of the $\Sigma$s derived from eqs.\,(4) and (10).

Figure \ref{Zcomps}a therefore represents the discussion in \S\ref{coplots} about the apparent consistency, but more subtle inconsistency, in the mass columns as measured by \hcop\ and \tco.  That discussion is equivalent to saying that the distribution of points in Figure \ref{Zcomps}a would be more self-consistent if, instead of straddling the diagonal dashed line of equality, the points were to lie below or to the right of it.  Although a suitable raising of \xhcop\ (lowering of $\Sigma_{\rm HCO^+}$), or a lowering of the [\tco]/[\htwo] ratio (raising of $\Sigma_{\rm ^{12}CO}$) could achieve this, we focus here instead on the alternative \ico\ to \nhtwo\ conversions described in \S\ref{newx}, and embodied in Figure \ref{Zcomps}b.

\notetoeditor{}
\begin{figure*}[t]
(a)\hspace{-2mm}\includegraphics[angle=0,scale=0.44]{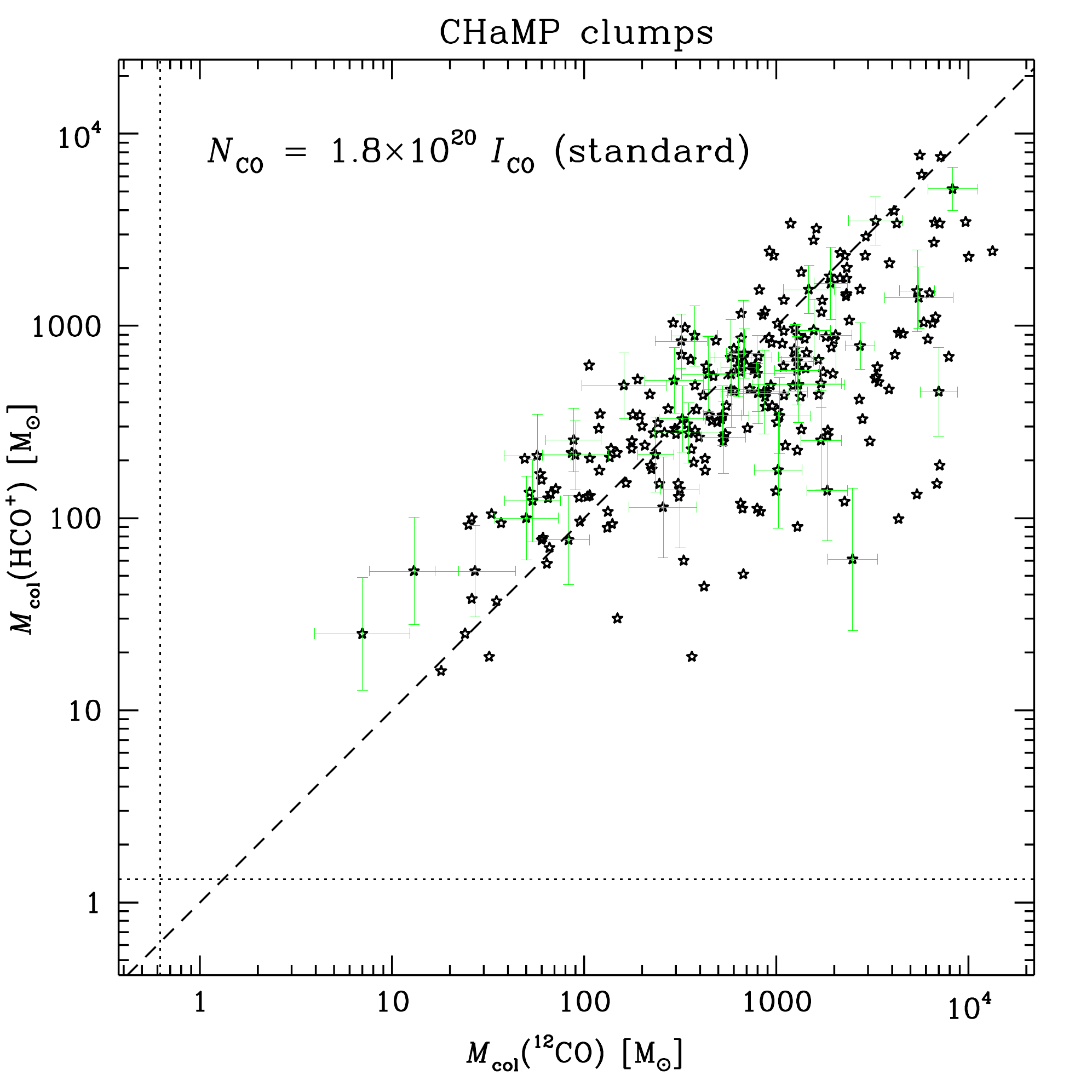}\hspace{2mm}
(b)\hspace{-2mm}\includegraphics[angle=0,scale=0.44]{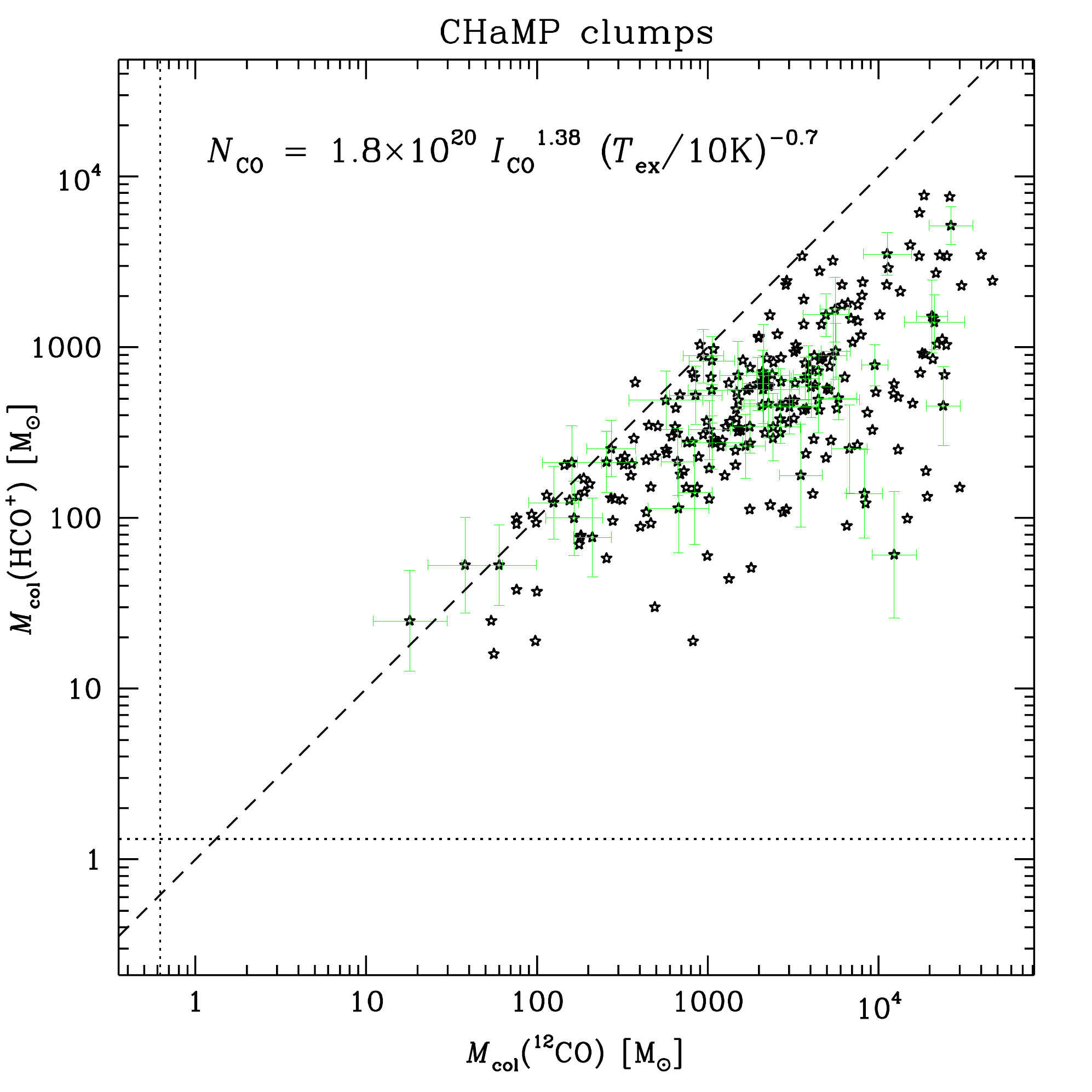}
\caption{Comparison of CHaMP clumps' masses derived from \hcop\ measurements in Paper I ($y$-axis in both panels, assuming \xhcop\ = 10$^{-9}$ as in Paper I) and that derived from \tco\ measurements presented here ($x$-axis in both panels, assuming [\tco]/[\htwo] = 10$^{-4}$).  The left panel (a) shows the implied $M$(\tco) using eq.\,(4), while the right panel (b) shows the same quantity but according to eq.\,(10), as labelled in each panel.  Other details are as in Fig.\,\ref{Zcomps}.
}
\label{Mcomps}
\end{figure*}

By using the alternative mass conversions, {\em which were derived for completely different samples of clouds}, and should therefore be free of any systematic bias in our data or methods, Figure \ref{Zcomps}b already brings the two $\Sigma$ estimates from the two CHaMP species into a more self-consistent distribution, namely to the right of the diagonal.  The overall shift in $\Sigma_{\rm ^{12}CO}$, from panel a to b in Figure \ref{Zcomps}, is by a factor of $\sim$3 higher, from an average $\Sigma$ ratio of $\sim$1 to $\sim$3.  This makes physical sense since, whatever the individual gas-phase abundances are of \hcop\ and \tco, if we are measuring the same discrete objects with our clump catalogue, we should find that the masses of the clumps including the less dense envelopes, should be at least the same as the masses of the denser interiors but not much more, given that the clump radii are only slightly larger and the gas density is dropping.

This result is surprisingly good, given the possibility that systematic effects could be embedded in the calculation at several points.  It seems unlikely to suppose that any such biases should exactly cancel to produce a result as ``clean'' as that in Figure \ref{Zcomps}b: only one point appears to the left of the diagonal in Figure \ref{Zcomps}b by more than the 1$\sigma$ uncertainty.  In contrast, Figure \ref{Zcomps}b would be less ideal if we did not include the \tex-dependence from the CMC project \citep[eq.\,9;][]{kl15}.  With the ThrUMMS result alone \citep[eq.\,8;][]{bm15}, the highest-$\Sigma$ points stretch much further to the right, and well away from the diagonal, to values ($>$10$^4$\,M\solar/pc$^2$) that seem unphysically large for the clumps involved.

A similar pattern is seen in the number density distributions, shown in Figure \ref{ncomps}, except there the change from panel $a$ to $b$ is less pronounced, since the slightly larger measured clump sizes in \tco\ act to counterbalance the increased mass column effected by using eq.\,(10).  Thus, the mean number density ratio has increased by a factor of $\sim$2.5, from 0.9 in panel $a$, to 2.3 in panel $b$.  This can be understood as follows: applying the eq.\,(10) conversion channel-by-channel, we obtain \nco\ cubes, despite the high \tco\ optical depths.  We then apply the procedure from Paper I to compute the peak density, which assumes a gaussian density profile in the cloud.

The clump mass distributions using the two conversion formulae are shown in Figure \ref{Mcomps}.  Again, we see that Figure \ref{Mcomps}a has some clumps with {\em total} masses (measured with \tco) {\em smaller} than that of their interiors (measured with \hcop), an unphysical result.  Altering \xhcop\ to bring panel $a$ of Figure \ref{Mcomps} to self-consistency would require a factor of 3--4 higher \hcop\ abundance, which in the mean seems unlikely, as explained in Paper I.  Panel $b$, on the other hand, resolves this discrepancy very cleanly, with {\em no points left of the diagonal} by more than 1$\sigma$.  The mean change in switching from eq.\,(4) to (10) is now a factor of $\sim$2 for the lower-mass clumps (which also have similar measured sizes in the two species) to $\sim$3--4$\times$ higher for the higher-mass clumps (of which a higher fraction have sizes much larger in \tco\ than \hcop, resulting in even higher integrated masses).  Indeed, in panel $b$, we see that the \tco-derived clump masses can range from very similar to the \hcop-derived masses, to 10$\times$ or 100$\times$ larger, meaning that the molecular mass fraction traced by \hcop\ may be quite small in some cases.  This is not unexpected, given that (1) some of our catalogued clumps will be on the threshold of sufficient density or column density to excite \hcop\ emission at all, and (2) we have deliberately omitted the many \tco\ clumps visible in our data which have {\em no} detectable \hcop.

\notetoeditor{}
\begin{figure*}[t]
(a)\hspace{-2mm}\includegraphics[angle=0,scale=0.44]{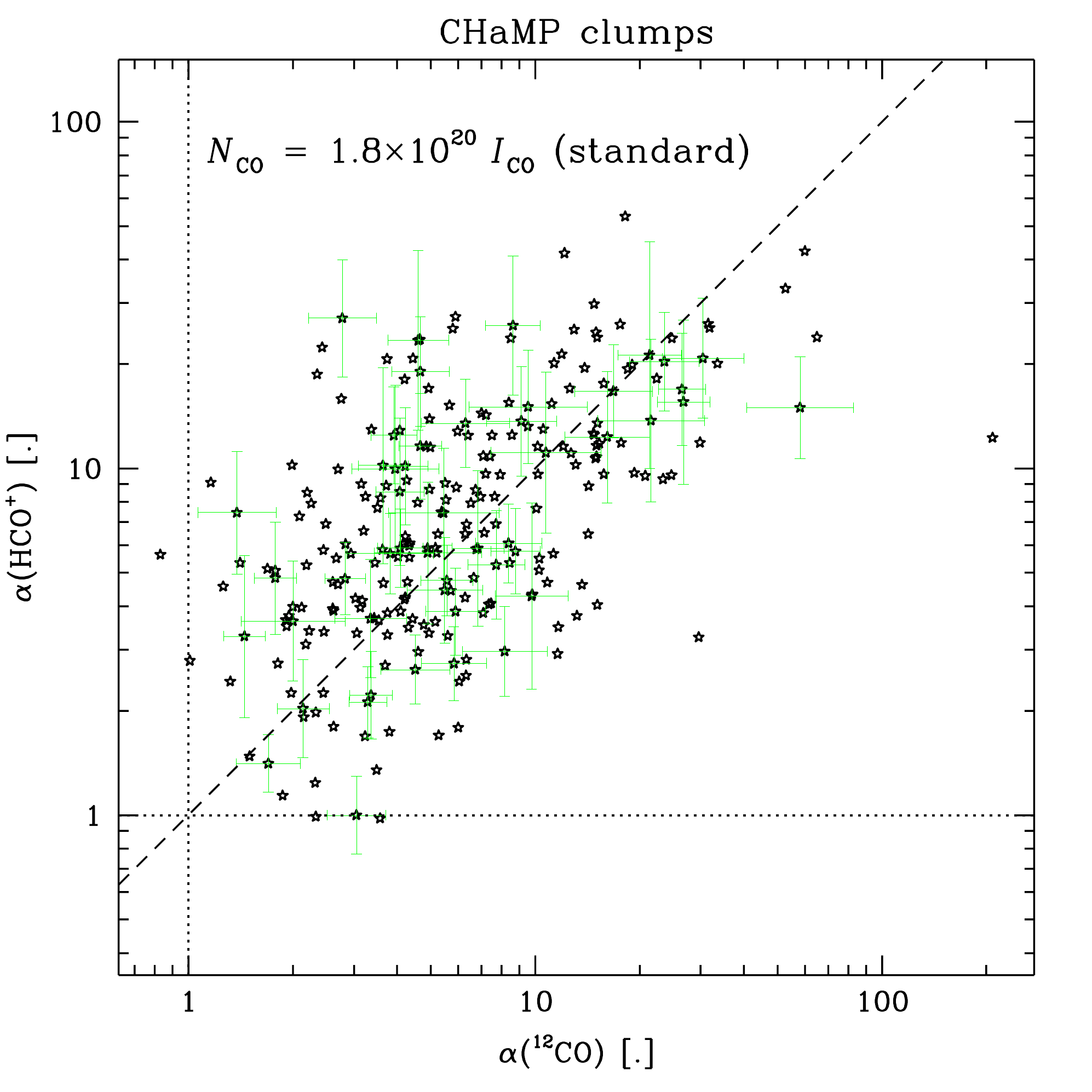}\hspace{2mm}
(b)\hspace{-2mm}\includegraphics[angle=0,scale=0.44]{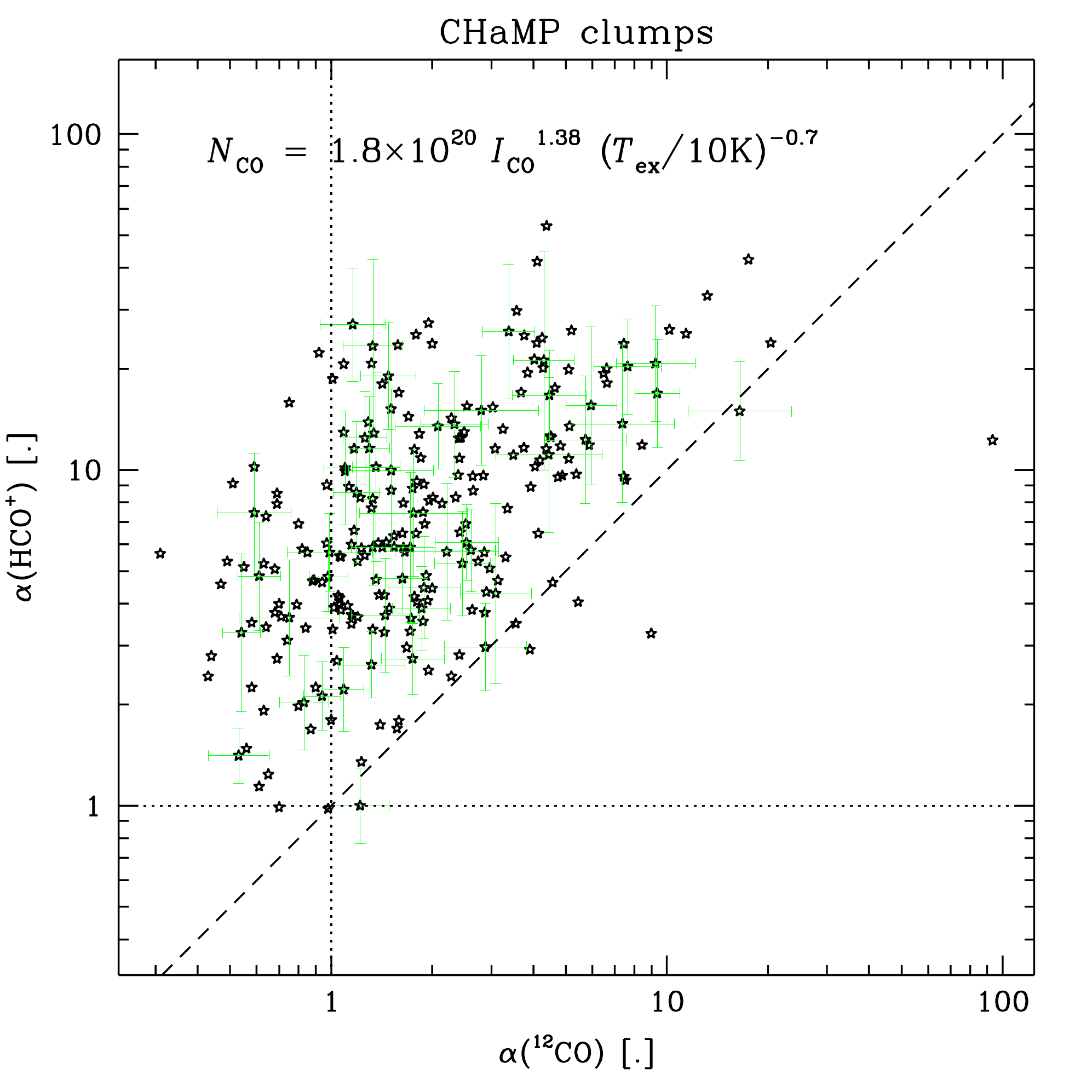}
\caption{Comparison of CHaMP clumps' virial-$\alpha$ derived from \hcop\ measurements in Paper I ($y$-axis in both panels, assuming \xhcop\ = 10$^{-9}$ as in Paper I) and that derived from \tco\ measurements presented here ($x$-axis in both panels, assuming [\tco]/[\htwo] = 10$^{-4}$).  The left panel (a) shows the implied $\alpha$(\tco) using eq.\,(4), while the right panel (b) shows the same quantity but according to eq.\,(10), as labelled in each panel.  Other details are as in Fig.\,\ref{Zcomps}.
}
\label{acomps}
\end{figure*}

This last point prompts an interesting corollary.  We can compare the total mass from the measured \tco\ clumps (Fig.\,\ref{Mcomps}, or Tables B2 or B3) to the total mass inferred from the \tco\ emission in {\bf{\em all our maps}}.  In the case of the standard $X$-factor (eq.\,4), the procedure is straightforward since the \ico\ to \nhtwo\ conversion is linear.  The total clump mass from Table B2 is 4.5$\times$10$^5$\,M\solar, which is 68\% of the mass obtained from integrating across all our maps, 6.7$\times$10$^5$\,M\solar.  If we consider each map separately, the mean $\pm$ SD of this mass ratio is $\Xi$ = 72$\pm$34\%, %
which is the {\em clump mass fraction} in the ``GMCs'' defined by our maps (most of the CHaMP maps contain more than the canonical GMC threshold of 10$^4$ M\solar\ by any measure of mass).  Therefore, there is some variation in $\Xi$ between GMCs, but overall the fraction is significantly higher than the dense core mass fraction of a few percent \citep{ll03,bh14}.  This nominal disparity depends strongly, however, on exactly what is meant by ``dense gas,'' a point to which we return below.

To do the same calculation with the new conversion formulae eqs.\,(8--10), we must proceed with caution since these are nonlinear in \ico.  This means we can't simply integrate the \tco\ emission in a given map and apply the conversion to the single integrated number.  Instead, we rewrite eq.\,(10) for the mass surface density (as eq.\,5 is derived from eq.\,4):
\begin{equation}   
	\Sigma_{\rm mol} = 3.38\,{\rm M}_{\odot}{\rm pc}^{-2}~\frac{(I_{\rm CO}/{\rm K\,km\,s}^{-1})^{1.38}}{(T_{\rm ex}/10\,K)^{0.7}}~,
\end{equation}
where \tex\ is taken from the peak \tco\ measurement via eq.\,(3).  With this we can convert our \tco\ cubes into equivalent cubes of mass surface density, and only then integrate the latter to obtain a total molecular mass for each cube.  (This was how the parameters in Table B3 were derived, per clump.)  Then the total mass of the clumps from Table B3 is 1.6$\times$10$^6$\,M\solar, while the integrated mass surface density cubes yield a total mass of 2.3$\times$10$^6$\,M\solar, or $\Xi$ = 72\%.  Evaluating each map separately, the mean $\pm$ SD for $\Xi$ is 78$\pm$38\%.  %
As expected given the nonlinearity of eq.\,(11) in \ico, these fractions $\Xi$ are $\sim$5\% higher than the ones obtained from eq.\,(5).

It is important to qualify this result with the recognition that it is a function of the mean gas density sampled by our clump population.  That is, it should be self-evident that the fraction of the molecular clouds' mass $\xi_p$ that lies above a given gas density $n=10^p$\,m$^{-3}$ will fall as $n$, or $p$, rises.  In the literature, however, the concept of ``dense gas'' has often been defined as a matter of convenience, i.e. as that gas traced by species like HCN, CS, \nht, etc.\,\citep[e.g.,][and references therein]{e99}.  Then the gas that is sampled in observational studies using these tracers is necessarily that gas near $n\sim10^{10}$\,m$^{-3}$, the effective density \citep{e99} of the low-$J$ transitions being mapped.  This is often the density that many workers refer to when using the term ``dense gas.''  However, most studies with these tracers are either much smaller in area or population size than ours, and are usually selected towards known dense cores in nearby star forming regions, and so are not unbiased.

\notetoeditor{}
\begin{figure*}[t]
(a)\hspace{-2mm}\includegraphics[angle=0,scale=0.44]{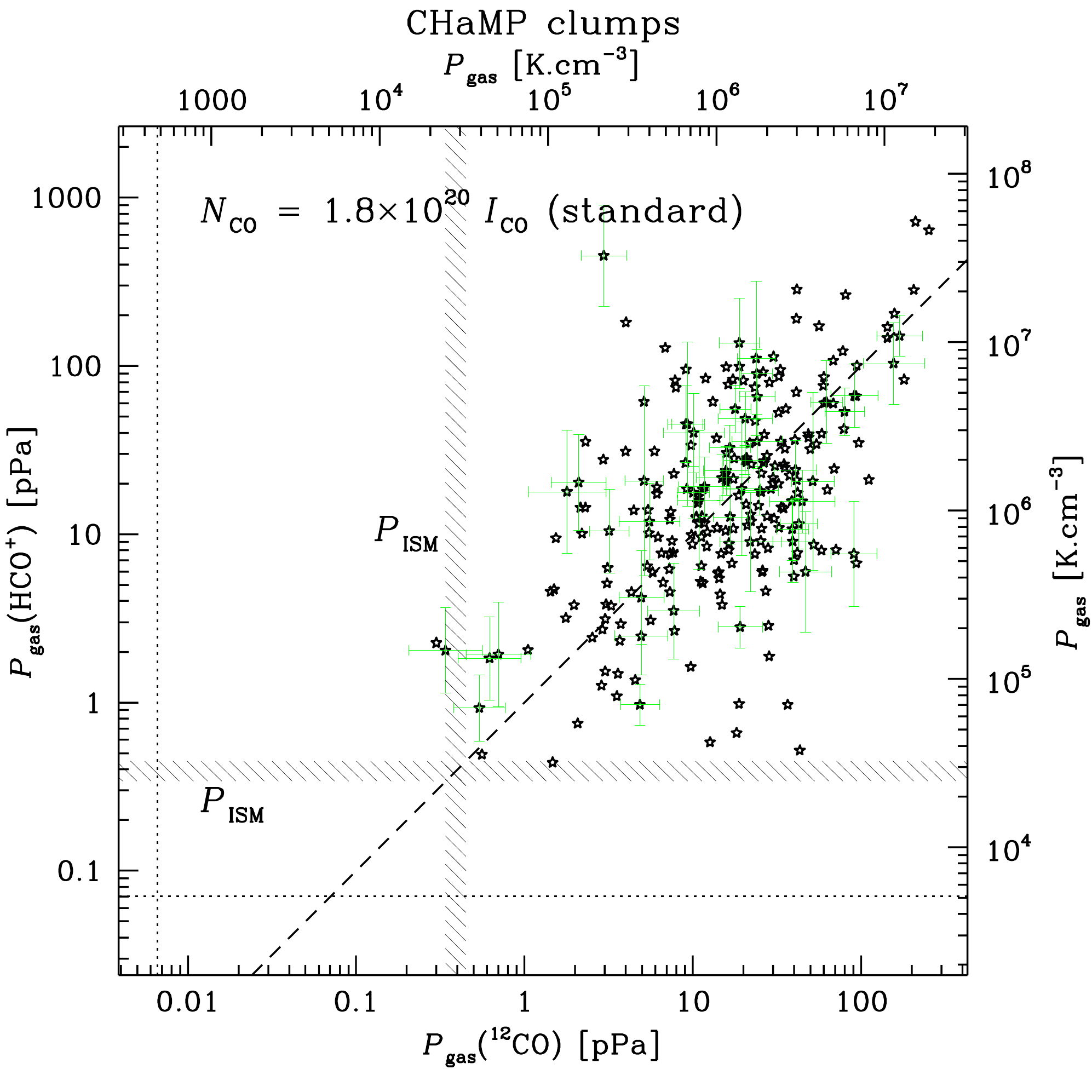}\hspace{2mm}
(b)\hspace{-2mm}\includegraphics[angle=0,scale=0.44]{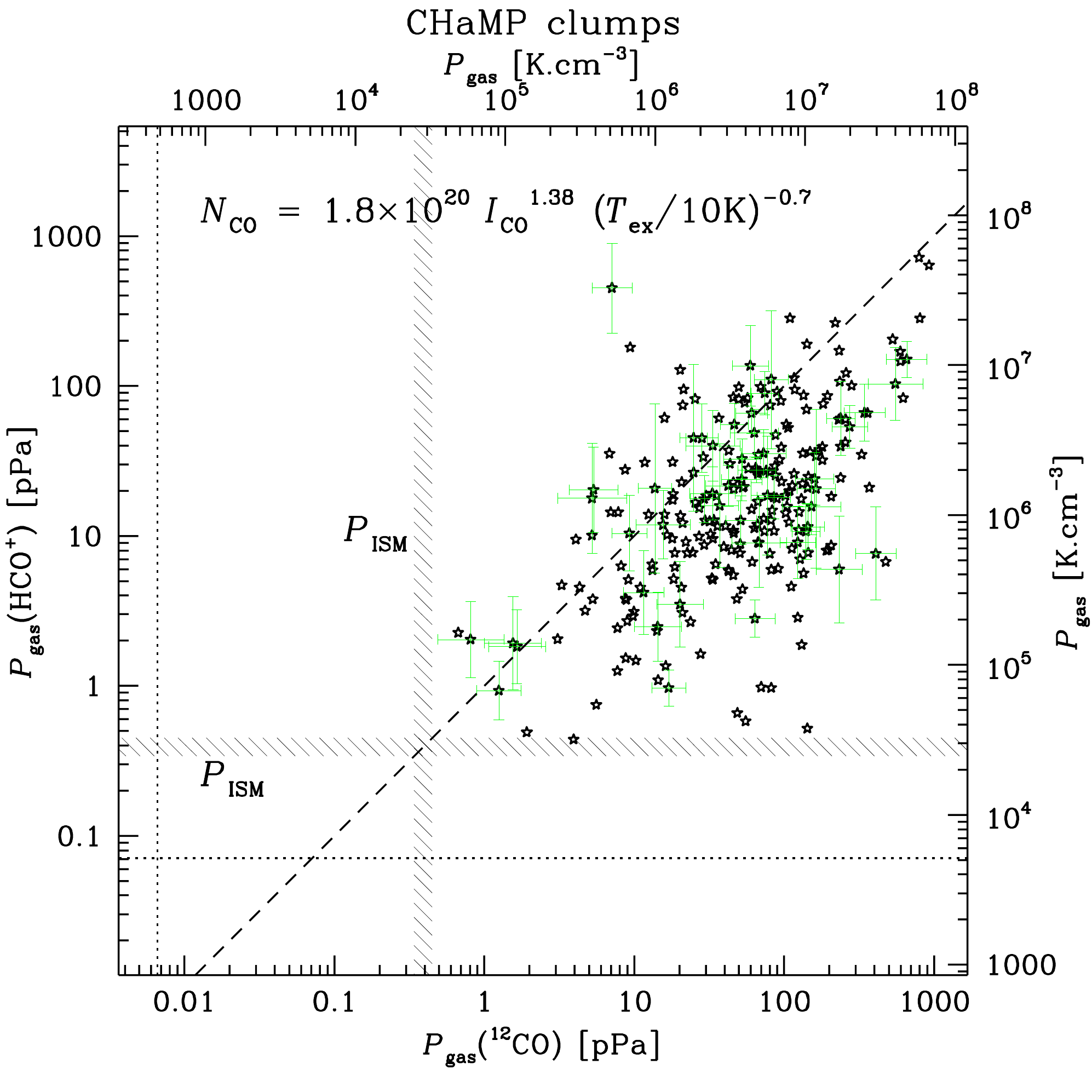}
\caption{Comparison of CHaMP clumps' internal pressure derived from \hcop\ measurements in Paper I ($y$-axis in both panels, assuming \xhcop\ = 10$^{-9}$ as in Paper I) and that derived from \tco\ measurements presented here ($x$-axis in both panels, assuming [\tco]/[\htwo] = 10$^{-4}$).  The left panel (a) shows the implied $P$(\tco) using eq.\,(4), while the right panel (b) shows the same quantity but according to eq.\,(10), as labelled in each panel.  Other details are as in Fig.\,\ref{Zcomps}.
}
\label{Pcomps}
\end{figure*}

As alluded to already, our use of \hcop, with an effective density 5--10$\times$ smaller than the above tracers, is more inclusive than this.  In Paper I, we showed that the denser interiors of our large, unbiased clump sample have typical peak densities $\sim$10$^9$\,m$^{-3}$, as would be expected.  Of these, a relatively small fraction ($\sim$2\%) have peak densities above 10$^{10}$\,m$^{-3}$.  Even with our revised conversion to \nhtwo\ described above, this fraction only rises to $\sim$7\%.  Understanding that these are the ``peak'' or central clump densities, the overall mass fraction $\xi_{10}$ (above 10$^{10}$\,m$^{-3}$) in our clumps will actually be less than these ratios, even while $\xi_9$ (the mass fraction above 10$^9$\,m$^{-3}$) is $\sim$50\% according to Fig.\,\ref{ncomps}a (and allowing for the fact that these are central densities), and $\xi_9\sim$ 90\% according to Fig.\,\ref{ncomps}b.  This is consistent with studies such as \citet{bh14} which find $\xi_{10}<$ 10\%.

The implication of these estimates is striking: not only do the low-column clumps omitted from this analysis but included in our maps (i.e., mapped \tco\ clumps without \hcop\ emission) {\bf plus} all the {\em extended} emission from the clumps analysed herein contribute a small fraction ($\sim$25\%) of the total molecular mass across all our maps, but more significantly, the very concept of a GMC is called into question.  The values of $\Xi$ approaching unity tell us that what we have traditionally called GMCs, including all molecular clouds containing dense gas, from the smaller variety (a few pc) to the largest GMCs (such as the $\eta$ Car cloud with an overall length of 120\,pc), are essentially just collections of parsec-scale clumps in terms of their overall structure and mass distribution.  Moreover, $\Xi$ shows no obvious trend with (e.g.) Region mass or size, and so its large value seems to be a general property of Galactic molecular clouds.

Changes to the virial-$\alpha$ results from using eq.\,(10) are also striking, and carry other important implications.  We show in Figure \ref{acomps} different $\alpha$ measurements in the same format as above.  Panel $a$ suggests that, as a whole (i.e., including their envelopes), some clumps may be considered more bound by gravity compared to just their interiors ($\alpha_{\rm ^{12}CO}$ $<$ $\alpha_{\rm HCO^+}$, those points to the left of the diagonal), while others are less bound at their envelopes than in their interiors ($\alpha_{\rm ^{12}CO}$ $>$ $\alpha_{\rm HCO^+}$, those points to the right of the diagonal).  Such a result would suggest that many clumps (the $\sim$half to the right of the diagonal) are truly unbound, since if the envelopes cannot provide pressure containment (i.e., a surface term in the Virial Theorem) to the dense-gas clumps as postulated by \citet{bm92}, then it is not clear that anything else can.  Even so, interiors that are ``more bound'' by the envelopes (points to the left of the diagonal) still have their $\alpha$ values mostly $\gg$1, suggesting that they, too, are not completely pressure-bound.  This would imply that the clump population as a whole is likely to be transient, which would be one explanation for the low overall star formation efficiency (SFE) of molecular clouds in the Milky Way, as discussed in Paper I.

In contrast, Figure \ref{acomps}b and eq.\,(10) suggest that {\em virtually all clumps} traced by \hcop\ are (at least partially) pressure-confined by their more massive envelopes, since nearly all of them have $\alpha_{\rm ^{12}CO}$ $<$ $\alpha_{\rm HCO^+}$, with $<$$\alpha$$>$ = 1.9.  Furthermore, 18\% (or 50 of 273 clumps in common) have $\alpha$ $<$ 1, and slightly more than 50\% of the clumps have $\alpha$ $<$ 2, suggesting that they are either {\em completely gravitationally bound} or near-virial-equilibrium structures.  This in turn supports the argument in Paper I, and also made from a different line of evidence by \citet{b13}, that the dense-gas clumps traced by \hcop\ are long-lived structures, which take time to accumulate sufficient density to engage in more than low levels of star formation for most of their lifetimes.

We therefore see that the use of the appropriate \ico\ to \nhtwo\ conversion is a rather critical component of the analysis of \tco\ emission from massive molecular clump samples such as ours, since the interpretation of the clumps' dynamical state depends strongly on getting this calculation right.  So while the ThrUMMS law embodies a numerically minor revision to the $X$-factor approach, physically and logically the improvement is significant.  Indeed, we come to a similar conclusion as \citet{kpg13}, that clump $\alpha$s are $\sim$2 on average, but with an intuitive physical explanation based on long-lived pressure confinement plus the new conversion law.

Thus, part of our motivation for presenting an $X$-factor analysis is to show that it does not give reasonable results for an optically thick line.  Consider virialisation of the clouds: for \tco, one would think that they are likely to have broader linewidths than in an optically thin transition.  This would mean that the virial masses and $\alpha$s derived from them are too big also.  However, Fig.\,\ref{RDVcomps}b shows that this is not the case.  So the reason that the $\alpha$s in Fig.\,\ref{alpha} are ``too large'' cannot be simply because of line broadening caused by high optical depth.  Instead, the ThrUMMS conversion shows that the $\alpha$s are too large because of the large optical depths themselves.  Thus, the lack of calibration of low-latitude $X$ values (i.e., prior to the ThrUMMS results) gives an inaccurate $X$ conversion for massive, parsec-scale clumps, as extensively discussed by \citet{dht01}.

Concluding this comparison, we show in Figure \ref{Pcomps} the total internal pressure measurements of the clumps, as in Paper I.  In panel $a$, as in Figure \ref{acomps}a, the calculation suggests that the clumps are roughly evenly divided between those more highly pressurised by their envelopes than in their interiors (to the right of the diagonal), and those more highly pressurised in their interiors than by their envelopes (to the left of the diagonal).  In panel $b$, however, we see a strong shift towards a clump population which is pressurised by their envelopes, although perhaps half of the clumps could be said to have roughly equal pressurisation inside and out (i.e., those within 1$\sigma$ uncertainty of the diagonal).  This division is then consistent with $\alpha$ = 2 being the dividing line, as in Figure \ref{acomps}b, between clumps that are bound or unbound overall.

\subsection{Global Implications}\label{global}

The above results show that high-quality molecular emission-line data on a well-defined sample of molecular clouds, and the careful analysis of these data, in particular the exact \ico\ to \nhtwo\ conversion, can have a very powerful influence on how we interpret not only our data, but also on our understanding of global processes in molecular cloud evolution and star formation.  

For example, a widely-held view of CO as a tracer of ``lower-density'' molecular gas, compared to ``dense gas tracers'' like \hcop\ or HCN, is that these are literal truths, based on the excitation requirements of these species \citep[e.g., see][for a discussion on how such views can be misleading]{e99}.  However, as we have seen here, this simple microscopic picture can mask the physics of real clouds.  In Figure \ref{deltaPA}b, we see that clumps' \hcop\ integrated intensity lies at a fairly constant level of 10\% of the \tco\ integrated intensity, until we run out of \hcop\ signal from the cloud periphery.  This suggests that there is probably \hcop-dark (or at least, \hcop-faint) gas in clump envelopes, and that both species must really coexist in the same volumes as demanded by astrochemistry.  This understanding then {\em requires} that the mass traced by \tco\ emission from clumps must be larger than that traced by \hcop.  If the clumps' internal density distribution can be approximated by a gaussian profile, then this also implies that the peak of the profile, as measured by the \tco\ emission, must be larger than the peak derived from the \hcop\ emission alone, as seen in Figure \ref{ncomps}b.  This is the {\em opposite} picture to that given by the microscopic view of molecular excitation.%

In this way, eq.\,(10) gives a much more ``satisfactory'' set of physical results in this study than eq.\,(4), or eqs.\,(8) or (9) alone.  However, this result does not by itself mean that eq.\,(10) is more correct than the other conversion prescriptions.  Eq.\,(4) also has much evidence in its favour and is widely-used \citep{hd15}, so the prospect of establishing that a new conversion like eq.\,(10) might be superior to eq.\,(4) would take more evidence than we can present here.  But it is very suggestive that the CHaMP data combined with eq.\,(10) --- the latter derived from two completely independent data sets \citep{kl15,bm15} --- can support, without any ``forcing'' of the results, a consistent picture previously proposed from disparate lines of evidence \citep{b11,b13}.

This picture centres around a long, quiescent lifetime for ``dense-gas'' (i.e., \hcop-bearing, or \hcop-bright) massive molecular clumps, where activity during this period may be limited to a low level of low-mass star formation, as first suggested by \citet{b11}, and similar to the long lifetime scenario of \citet{kss09}, based on the molecular cloud population of M51.  The results of the work presented here suggest that this quiescent phase is {\em enabled by a pressure- and gravity-confined massive molecular envelope}, that stabilises a less massive and still turbulent dense interior, which by itself is not dense enough to be gravitationally bound and efficiently form a star cluster.  Only once a certain density threshold is crossed (perhaps $\sim$10$^{10}$\,m$^{-3}$ according to the results of Paper I), does the SFE of the clump's dense interior rise to the point that vigorous, massive star and star-cluster formation can take place.  If this latter phase takes a few $\times$ 10$^6$\,yr, then the clump statistics would imply that the quiescent phase takes $\sim$6--20$\times$ longer, yielding overall clump lifetimes that might range over 20--100 $\times$ 10$^6$\,yr.

We are continuing with other work to investigate whether this picture holds, or the standard column density conversion (eq.\,4) holds and the clumps we see represent a more ephemeral population of clouds which are constantly forming and dissipating.  Data collected from other studies for other purposes may also be re-analysed to examine such questions.  For example, eq.\,(10) also implies a somewhat different calibration for the Kennicutt-Schmidt relations.  Studies re-examining the KS relations may also find implications for long clump lifetimes, and consequently longer gas depletion timescales in disk galaxies, than have previously been discussed, supporting the hypothesis argued for here.

\section{Conclusions}

As the second major mm-wave data release of the CHaMP project, we have presented new observations and analysis of the \tco\ line emission from a complete sample of $\sim$300 massive molecular clumps, originally defined by their \hcop\ emission in Paper I.  The \tco\ emission traces the less dense molecular envelopes in which the denser \hcop-bearing interiors are embedded, and are the first results from Phase II of the Mopra mapping during 2009--12, which covered several molecular lines in the 107--115\,GHz range.

The \tco\ observing and data processing include three significant advances over the techniques utilised in Paper I for \hcop: 

\hspace{1mm} \put(5,3){\circle*{3}} \hspace{2mm} We have used a sky-adaptive version of the standard on-the-fly (OTF) mapping technique, called ``Active Mapping'' or AM, which adjusts the mapping speed of each map for the sky conditions (whether elevation- or weather-dependent) at the time of observation.  OTF-AM has the effect of making the rms noise levels in each map much more uniform than with regular OTF.

\hspace{3mm} \put(5,3){\circle*{3}} \hspace{2mm} We have used data-screening techniques to filter out errors and correct for calibration problems.

\hspace{3mm} \put(5,3){\circle*{3}} \hspace{2mm} We have used a smooth-and-mask (SAM) technique to dramatically improve the image fidelity in the moment maps derived from the data cubes, even for the intrinsically high S/N \tco\ data.

With these new data and techniques, we have compiled the observed and derived physical properties of the identified ``massive dense clump'' envelopes, and compared these properties with those of the clump interiors as presented in Paper I.  Our main results are as follows:

\vspace{0mm}1.\ The observed clump sizes and linewidths are very similar in both \tco\ and \hcop, with only a $\sim$25\% contribution to the total mass of GMCs and molecular cloud complexes from extended ($>$5\,pc) cloud components.

\vspace{0mm}2.\ This suggests that parsec-scale clumps ($<$$R$$>$ = 0.84\,pc, with a logarithmic dispersion equivalent to a factor of 1.9 around this size) comprise the basic building blocks 
of the molecular ISM, and that in the main, extended molecular structures such as GMCs are just collections of such clumps, since most of the mass, $\Xi\sim$ 75\%, is enclosed by the parsec-scale clumps we sample, with typical central densities 10$^{9-9.5}$\,m$^{-3}$.

\vspace{0mm}3.\ We see a weak, but real, Larson-type size-linewidth relation for the envelopes, with $\sigma_{\rm V}$ = (1.67 $\pm$ 0.04\,\kms)\,$R_{\rm pc}^{0.24\pm0.04}$, whereas for \hcop\ we saw no statistically significant size-linewidth relation. 

\vspace{0mm}4.\ When computed using a standard $X$-factor, the whole-clump properties including the envelopes give slightly lower central volume densities, and similar column densities, masses, virial-$\alpha$, and total internal pressures compared to the interiors alone.

\vspace{0mm}5.\ When computed using new \ico\ to \nhtwo\ conversion formulae from \citet{bm15} and \citet{kl15}, the whole-clump properties including the envelopes give somewhat higher central volume densities and total internal pressures, systematically higher column densities and masses, and systematically lower virial-$\alpha$ compared to the interiors alone.

\vspace{0mm}6.\ We interpret these results to mean that, including the envelope mass, $\sim$half the clumps detected in \hcop\ are gravitationally bound or near virial equilibrium, even when their interiors alone are not.

\vspace{0mm}7.\ This suggests that about half the \hcop-defined ``dense-clump'' population are truly pressure-confined by their massive envelopes, as originally postulated by \citet{bm92}.

\vspace{0mm}8.\ This in turn supports the view that a significant fraction of the observed dense clumps are long-lived structures, apparently in a state of low star formation rate/efficiency over several tens of Myr, until accumulating sufficient matter to pass a density threshold of $\sim$10$^{10}$ \htwo\ molecules m$^{-3}$, and only then engaging in vigorous massive star formation at the end of this time.

\vspace{0mm}9.\ This is consistent with other studies that find a relatively small mass fraction $\xi_p<$ 10\% of molecular gas above a density $n=10^p$\,m$^{-3}$ for $p$=10, even while we find $\xi_p$ \gapp 50\% for $p$=9.  It seems to be this smaller fraction of denser ``dense gas'' which is actively engaged in star formation, as found in several other studies.

\vspace{0mm}We look forward to further tests of the scenario argued for here, and proposed by \citet{b11}, that the long lifetime for massive dense clumps' quiescent phase is enabled by a pressure- and gravity-confined massive molecular envelope, that stabilises a less massive and still turbulent dense interior, which by itself is not dense enough to be gravitationally bound and efficiently form a star cluster until a sufficient density threshold is passed.
\vspace{0.1mm}


\vspace{-8mm}
\acknowledgments

We thank Phil Edwards, Balt Indermuehle, and the ATNF staff for their support of the Mopra telescope.  PJB acknowledges support from NASA/JPL contract RSA-1464327, NSF grant AST-1312597, and the UF Astronomy Department.  AKH acknowledges support from grants awarded to B. Wakker at the University of Wisconsin-Madison.  SNO acknowledges support from the University of Florida Astronomy Department and the UF University Scholar's Program.  We also thank the anonymous referee for prompting several clarifications, and the discussion about $\xi$ and $\Xi$. %

Facilities: \facility{Mopra(MOPS)}.

\end{document}